\documentclass[11pt,twocolumn,twoside]{ATOM}
\usepackage{apacite}
\usepackage{natbib}
\usepackage{multirow}

\graphicspath{{./}{Figures/}}

\title{Original Research By Young Twinkle Students (ORBYTS): Ephemeris Refinement of Transiting Exoplanets III}

\author[1,2,3,$^\dagger$,*]{Billy Edwards}
\author[4,*]{Cynthia S. K. Ho}
\author[4,*]{Hannah L. M. Osborne}

\author[5,*]{Nabeeha Deen}
\author[5,*]{Ellie Hathorn}
\author[5,*]{Solomon Johnson}
\author[5,*]{Jiya Patel}
\author[5,*]{Varun Vogireddy}
\author[5,*]{Ansh Waddon}

\author[6,*]{Ayuub Ahmed}
\author[6,*]{Muhammad Bham}
\author[6,*]{Nathan Campbell}
\author[6,*]{Zahra Chummun}
\author[6,*]{Nicholas Crossley}
\author[6,*]{Reggie Dunsdon}
\author[6,*]{Robert Hayes}
\author[6,*]{Haroon Malik}
\author[6,*]{Frank Marsden}
\author[6,*]{Lois Mayfield}
\author[6,*]{Liston Mitchell}
\author[6,*]{Agnes Prosser}
\author[6,*]{Valentina Rabrenovic}
\author[6,*]{Emma Smith}
\author[6,*]{Rico Thomas}

\author[1]{Anastasia Kokori}
\author[1]{Angelos Tsiaras}
\author[2,1]{Marcell Tessenyi}
\author[1,2]{Giovanna Tinetti}
\author[1,2]{Jonathan Tennyson}

\affil[1]{Department of Physics and Astronomy, University College London, Gower Street, London, WC1E 6BT, United Kingdom}
\affil[2]{Blue Skies Space Ltd., 69 Wilson Street, London, EC2A 2BB, United Kingdom}
\affil[3]{AIM, CEA, CNRS, Universit\'e Paris-Saclay, Universit\'e de Paris, F-91191 Gif-sur-Yvette, France}
\affil[4]{Mullard Space Science Laboratory, University College London, Holmbury St. Mary, Dorking, RH5 6NT, United Kingdom}
\affil[5]{London Academy of Excellence, 322 High Street, London, E15 1AJ, United Kingdom}
\affil[6]{Highams Park School, Handsworth Avenue, London, E4 9PJ, United Kingdom}

\affil[$^\dagger$]{Corresponding author: billy.edwards.16@ucl.ac.uk}
\affil[*]{These authors contributed equally to this work.}

\dates{Received \today}


\doi{\url{https://doi.org/10.32374/atom.2020.2.4}} 

\begin{abstract}

We report photometric follow-up observations of thirteen exoplanets (HATS-1\,b, HATS-2\,b, HATS-3\,b, HAT-P-18\,b, HAT-P-27\,b, HAT-P-30\,b, HAT-P-55\,b, KELT-4A\,b, WASP-25\,b, WASP-42\,b, WASP-57\,b, WASP-61\,b and WASP-123\,b), as part of the Original Research By Young Twinkle Students (ORBYTS) programme. All these planets are potentially viable targets for atmospheric characterisation and our data, which were taken using the LCOGT network of ground-based telescopes, will be combined with observations from other users of ExoClock to ensure that the transit times of these planets continue to be well-known, far into the future. 

\end{abstract}

\setboolean{displaycopyright}{true}

\begin{document}

\maketitle

\thispagestyle{fancy}
\ifthenelse{\boolean{shortarticle}}{\abscontent}{}


\section{Introduction}

\phantomsection

\subsection{A Brief History of Exoplanet Detections} 

The concept of extrasolar planets, worlds which orbit other stars, has existed since at least the era of the Ancient Greeks, with Democritus and Epicurus believing in the idea that there existed an infinite amount of worlds, some of which possessed their own organisms. Much later, Italian cosmological theorist Giordano Bruno encouraged the idea that the countless stars were other suns which could each host planets similar to those in our own Solar System. Additionally, he reinforced the idea that there are other inhabited worlds that exist in the universe. After Neptune was mathematically predicted \citep{le_verrier_neptune} and then successfully detected \citep{galle_neptune}, attention turned to applying these methods to searching for planets around other stars.

A number of different detection techniques were devised and several detection claims were made \citep{jacob_1855,see_1896,van_kemp}, each of which in turn was shown to be spurious \citep{sherrill,boss}. The effectiveness of several of the detection techniques were underestimated due to the expectation that other planetary systems would resemble our own. While many early efforts focused on astrometry, studying the perturbations in the positions of stars, the radial velocity method, which uses Doppler shifts in the star light to infer the presence of a companion, delivered the first detection of a planetary body around a main-sequence star: 51 Pegasi b \citep{mayor_51peg}. Several years prior to this, a planetary system had been detected around a pulsar via precise timing measurements of the pulses \citep{wolszczan_pulsar} and a previous potential detection of a planetary body around Gamma Cephei \citep{campbell_1988} was later verified \citep{hatzes_2003}. 

After the discovery of 51 Pegasi b, numerous other detections were made using the radial velocity method, with many of these being worlds which are now referred to as hot Jupiters, large planets that have orbital periods of less than around 10 days. Notably, it had been argued more than 40 years earlier that, with the best spectrographs available at the time, such planets could be detected by the radial velocity method \citep{struve_1952}. While the radial velocity method can provide the mass and period of the planet, the radius cannot be determined. However, if the geometry of the system is aligned in the correct way, the planet can be seen to pass between the observer and its host star (a planetary \textit{transit}). The decrease in the flux is dependent upon the ratio of the planet's and the star's radii and thus provides a key planetary characteristic. Additionally, when combined with radial velocity data a constraint on the density can be placed and, in some cases, the bulk composition inferred. Therefore, many studies searched for transits of these planets with HD\,209458\,b being the first planet to be seen to occult its host star \citep{Charbonneau_2000,Henry_2000}.

\subsection{Current Status}

Since these early detections, there has been a rapid increase in the number of known exoplanets, with over 4400 having been identified by September 2021. While many different methods have been successfully used to detect planets, the most lucrative thus far has been the transit technique, with a number of ground-based and space-based surveys contributing to this deluge of detections. Indeed, of the exoplanets discovered to date, around 75.5\% have been detected using the transit method\footnote{\url{https://exoplanetarchive.ipac.caltech.edu/docs/counts\_detail.html}}. The Kepler space telescope is perhaps the most famous and influential transit survey, contributing more than 2500 planets and many more candidates \citep{borucki}. The most recent major exoplanet discovery mission to be launched is the Transiting Exoplanet Survey Satellite (TESS). Beginning operations in mid-2018, this mission is surveying hundreds of thousands stars across the entire sky \citep{ricker} and has already been successful in finding nearly 4400 candidate signals as well as confirming the existence of over 120 exoplanets\footnote{\url{https://tess.mit.edu/publications/}}.

Many of the planets found by TESS will be around bright stars, making them amenable for further characterisation. While current ground-based and space-based facilities have begun characterising the atmospheres of a handful of exoplanets, it is the next generation of facilities that offer the opportunity to truly move into an era of characterisation. The future of space-based facilities is especially promising and the immanent launch of the James Webb Space Telescope (JWST) is eagerly anticipated, with several programmes dedicated to studying transiting exoplanets \citep[e.g.][]{bean_ers}. Furthermore, Twinkle\footnote{\url{https://www.twinkle-spacemission.co.uk/}}, an upcoming, 0.45\,m space-based telescope, will conduct a dedicated extrasolar survey which will include the characterisation of many exoplanetary atmospheres \citep{twinkle}. Finally, Ariel is the M4 mission in ESA's Cosmic Vision programme which is scheduled to launch in 2029. Ariel will investigate the atmospheres of over 1000 transiting exoplanets using visible and near-infrared spectroscopy \citep{tinetti_ariel,tinetti_ariel2}.

These three facilities, in addition to continued observations from the ground and with Hubble, offer a golden future for characterising transiting exoplanets. However, the next generation of telescopes will require rigorous scheduling to minimise overheads and maximise science outputs. As such, interesting science targets could see their observing priority degraded if their ephemerides are not accurate enough, even if they are excellent targets for atmospheric characterisation. Many currently known planets have large ephemeris uncertainties and analysis suggests many TESS targets will have errors of $>$30 minutes less than a year after discovery due to the short baseline of TESS observations \citep{dragomir}. Therefore, detections by this mission, as well as other transiting planets, will have to be regularly followed-up to ensure their ephemerides remain well-known \citep{kokori_exoclock,zellem_exowatch}.


\subsection{Aims of the Project}

Our project was undertaken as part of the Original Research By Young Twinkle Students (ORBYTS) programme, which unites academic researchers with secondary school students. As part of the programme, pupils work on original research linked to space science \citep{sousa_silva_orbyts}. Since the programme’s foundation in 2016, over 150 school students have published research in academic journals through ORBYTS. The topic of this research has varied, from calculating empirical molecular energy levels \citep{mckemmish_orbyts,mckemmish_orbytsII,chubb_orbytsI,chubb_orbytsII,darby_orbyts}, analysing data of our Sun from the Hinode spacecraft \citep{french_orbyts} or studying protostellar outflows \citep{holdship_orbyts}, to counting craters on Mars \citep{francis_orbyts}, monitoring X-rays from Jupiter's Auroras \citep{wibisono_orbyts} and active galactic nuclei \citep{grafton_orbyts}.

In this project, we continued the work of previous ORBYTS groups \citep{edwards_orbytsII,edwards_orbytsI} in aiming to observe the transits of extrasolar planets which are suitable for atmospheric characterisation. By doing so, we aim to help ensure the planets' ephemerides will be well-known such that future facilities can characterise the atmospheres of these planets.

\section{Methods}

\begin{figure}
    \centering
    \includegraphics[width=\columnwidth]{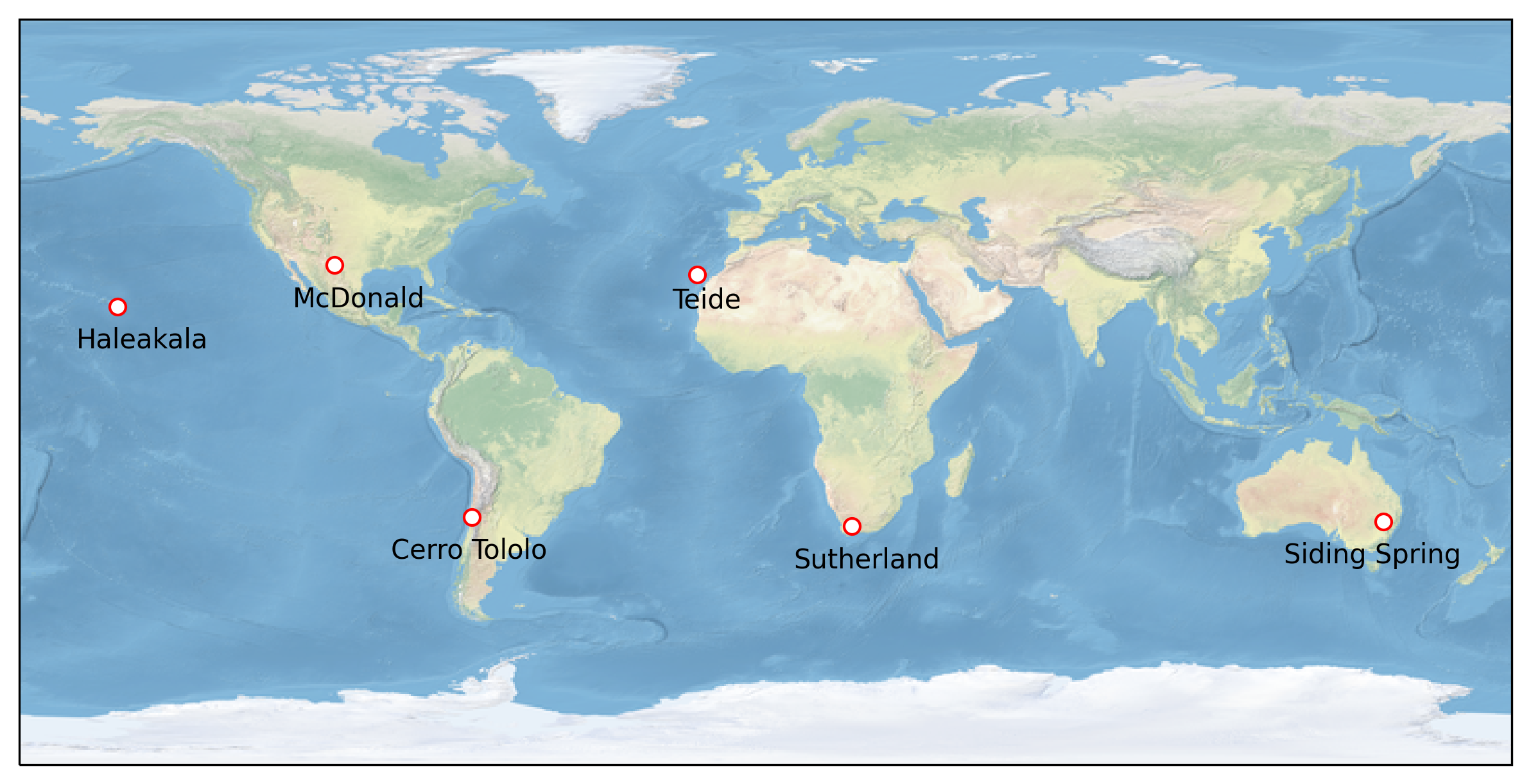}
    \caption{Locations of the LCOGT's network of robotic 0.4\,m telescopes.}
    \label{fig:lco_loc}
\end{figure}

We utilised the Las Cumbres Observatory Global Telescope (LCOGT) network's ground-based 0.4\,m telescopes \citep{brown_lco}, with access provided via the Global Sky Partners programme\footnote{\url{https://lco.global/education/partners/}} and the Faulkes Telescope Project\footnote{\url{http://www.faulkes-telescope.com/}}. The network has six sites which host 0.4\,m telescopes and these are spread across both the northern and southern hemispheres as shown in \autoref{fig:lco_loc}.

\phantomsection

\subsection{Target Selection and Data Collection}

\begin{figure}
    \centering
    \includegraphics[width=\columnwidth]{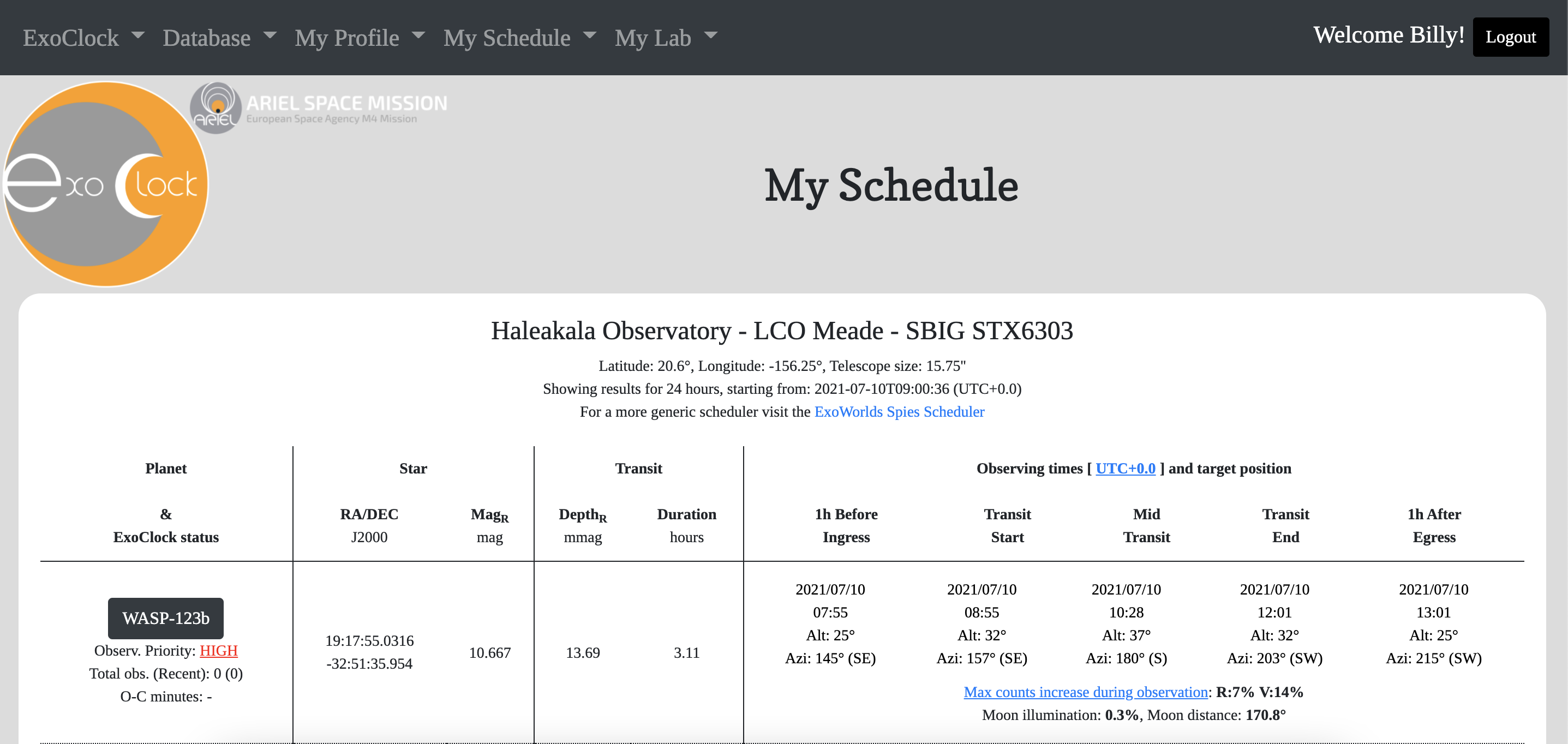}
    \caption{Screenshot of the ExoClock schedule tool, which provided a list of potential transit observations that could be conducted with the telescopes listed under our account.}
    \label{fig:exoclock}
\end{figure}

We used ExoClock\footnote{\url{https://www.exoclock.space/}} \citep{kokori_exoclock} to prioritise targets for ephemeris refinement. The site contains a database of all the exoplanets that could potentially be studied with Ariel \citep{ariel_targets}. These are ranked as low, medium or high priority based upon the current uncertainty on their transit times, the predicted precision in 2028, and the time since they were last observed. By loading in the size and location of your telescope(s), ExoClock provides a list of potential observations over the coming days. An example of this schedule is shown in \autoref{fig:exoclock} and, from the long list of potential planets to observe, we focused only on those ranked as medium or high priority.

\begin{figure}
    \centering
    \includegraphics[width=\columnwidth]{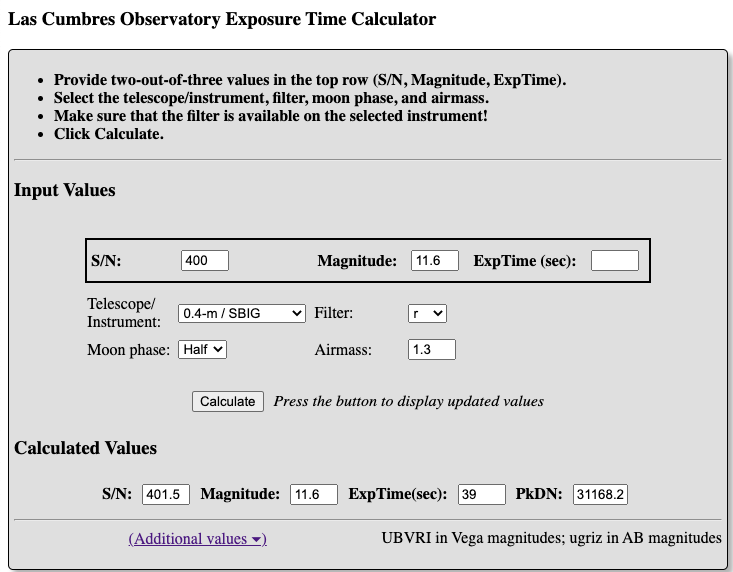}
    \caption{The LCO Exposure Time Calculator used to check the predicted quality of our observations and to ensure the telescope's detector didn't saturate.}
    \label{fig:exp_calc}
\end{figure}

Before using the LCO portal to book an observation of a potentially suitable target, we first calculated an exposure time using the LCO exposure time calculator\footnote{\url{https://exposure-time-calculator.lco.global/}}. To do this, we used the R-band magnitude of the host star from the ExoClock site and ensured the 0.4\,m telescope option was selected as shown in \autoref{fig:exp_calc}.


We calculated the required signal-to-noise ratio (SNR) for our observations from:
\begin{equation}
    \rm{SNR} > 5 \times \frac{1000}{\delta_R}
    \label{eq:snr}
\end{equation}
where $\delta_{\rm{R}}$ is the transit depth in mmag in the R-band which was taken from the ExoClock site. Having ensured that the required SNR could be reached without saturation, we booked our observations through the LCO portal. Due to issues with the weather or competing schedules, not all our observing requests were successful. However, we acquired data for thirteen planets: HATS-1\,b \citep{penev_h1}, HATS-2\,b \citep{mohler_h2}, HATS-3\,b \citep{bayliss_hats3}, HAT-P-18\,b \citep{hartman_hatp18}, HAT-P-27\,b \citep{anderson_h27,beky_h27}, HAT-P-30\,b \citep{johnson_h30}, HAT-P-55\,b \citep{juncher_h55}, KELT-4A\,b \citep{eastman_k4}, WASP-25\,b \citep{enoch_w25}, WASP-42\,b \citep{Lendl_WASP-42}, WASP-57\,b \citep{faedi_w57}, WASP-61\,b \citep{hellier_w61} and WASP-123\,b \citep{turner_w123}. 

\subsection{Data Reduction and Analysis}

We used the HOlomon Photometry Software (HOPS, \citep{tsiaras_hops}), which is freely available on GitHub\footnote{\url{https://github.com/ExoWorldsSpies/hops}}, to analyse the datasets we acquired. 

The first step of the analysis within HOPS is usually an initial reduction of the datasets (dark, flat and bias subtraction). However, LCO already performs the initial reduction for us: we obtained the reduced (BANZAI) data from the LCO archive and proceeded to process the data by uploading it onto HOPS. The filter was set to the R filter, which is the optical filter we used to observe all exoplanets in this study, and the co-ordinates of the host-star were obtained from the Right Ascension/Declination data found in the file's header. Within HOPS, we inspected frames for which sudden changes in the sky ratio or PSF were seen and any images that were deemed poor quality were removed. Due to the Earth's rotation, and slight errors in the telescope's ability to track the host star, the position of stars on the detector focal plane can change over the course of a night. Therefore, HOPS aligns the images to ensure the location of each star within the image is constant so the star's flux can be accurately measured over time.

Next, we extracted the flux from the target star. HOPS was simple to use in this regard, we only needed to pick our target star, and comparison stars to remove variations in the star's flux that were not due to the planet. Given that comparison stars may be variable, we inspected the photometry to ensure no spurious signals were being inserted into the host star's flux. If any comparison stars were deemed inappropriate, we removed and/or replaced them to achieve more stable light curves. 

\begin{table*}[]
    \centering
    \caption{Summary of observations undertaken as part of this project.}
    \begin{tabular}{cccccc}\hline\hline
     Planet & Star Mag [R] & Exposure Time [s] & Filter & Facility & Date\\\hline
    HATS-1\,b  & 12.08 & 60.284 & SDSS-rp & Cerro Tololo & 21/02/2021 \\ 
    HATS-2\,b  & 13.40 & 120.237 & SDSS-rp & Siding Spring & 22/02/2021\\ 
    HATS-3\,b  & 11.69 & 45.280 & SDSS-rp & Teide & 17/05/2021\\
    HAT-P-18\,b  & 12.61 & 90.285 & SDSS-rp & McDonald & 07/05/2021\\ 
    HAT-P-18\,b  & 12.61 & 90.288 & SDSS-rp & McDonald & 18/05/2021\\ 
    HAT-P-27\,b  & 11.98 & 59.937 & SDSS-rp & Haleakala & 18/04/2021 \\
    HAT-P-30\,b  & 10.104 & 20.281 & SDSS-rp & Teide & 03/02/2021 \\
    HAT-P-55\,b  & 12.87 & 61.939 & SDSS-rp & Haleakala & 16/05/2020 \\
    HAT-P-55\,b  & 12.87 & 69.945 & SDSS-rp & Haleakala & 31/03/2021 \\
    KELT-4A\,b  & 9.90 & 10.282 & SDSS-rp & Teide & 01/03/2021 \\ 
    WASP-25\,b  & 11.82 & 30.286 & SDSS-rp & Sutherland & 23/02/2021 \\
    WASP-42\,b  & 11.71 & 45.233 & SDSS-rp & Siding Spring & 17/05/2021\\ 
    WASP-57\,b  & 12.90 & 90.288 & SDSS-rp & Siding Spring & 21/04/2021 \\
    WASP-61\,b  & 11.88 & 60.285 & SDSS-rp & McDonald & 15/12/2020 \\
    WASP-123\,b  & 10.67 & 20.288 & SDSS-rp & Cerro Tololo & 28/07/2021 \\\hline\hline
    \end{tabular}
    \label{tab:observations}
\end{table*}

\begin{figure*}
    \centering
    \includegraphics[width=0.9\textwidth]{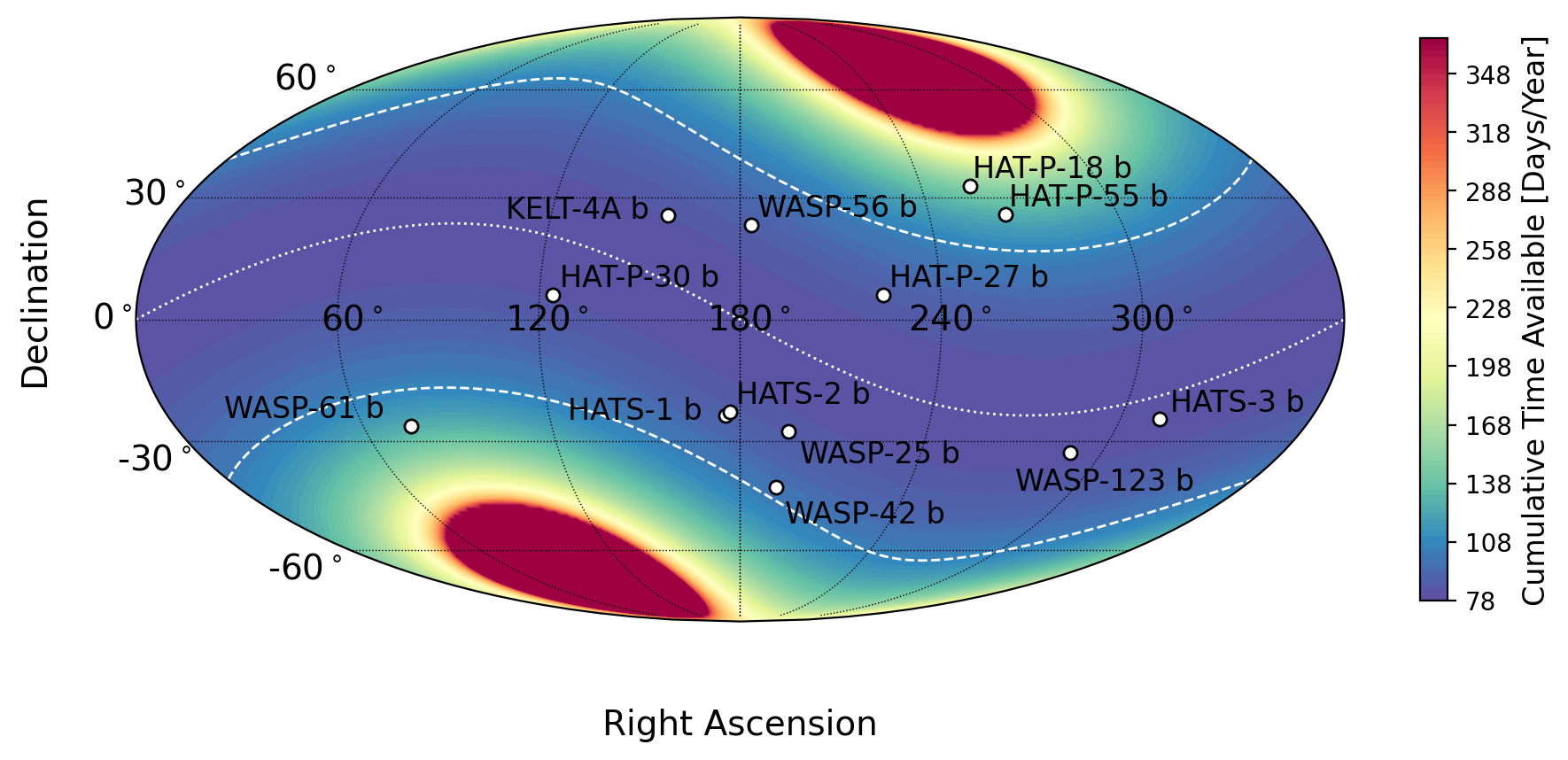}
    \caption{The sky coverage of Ariel, given in days available per year, with the planets studied in this paper over-plotted. The dotted white line shows the ecliptic plane while the dashed white lines represent the extent of Twinkle's field of regard, indicating that, of the planet studied in this paper, only HAT-P-18\,b, HAT-P-55\,b and WASP-61\,b are not also potential targets for this mission.}
    \label{fig:RA/DEC}
\end{figure*}

Finally, we used the built-in transit fitting feature of HOPS, which uses pylightcurve\footnote{\url{https://github.com/ucl-exoplanets/pylightcurve}} \citep{tsiaras_plc}, to fit our data. The parameters used for the fitting were those in the ExoClock database, which were in turn taken from the following papers: HATS-1\,b \citep{penev_h1}, HATS-2\,b \citep{mohler_h2}, HATS-3\,b \citep{bayliss_hats3}, HAT-P-18\,b \citep{seeliger_h18_h27}, HAT-P-27\,b \citep{seeliger_h18_h27}, HAT-P-30\,b \citep{mac_h30}, HAT-P-55\,b \citep{juncher_h55}, KELT-4A\,b \citep{eastman_k4}, WASP-25\,b \citep{southworth_w25}, WASP-42\,b \citep{southworth_w42}, WASP-57\,b \citep{Southworth_WASP-57}, WASP-61\,b \citep{hellier_w61}, WASP-123\,b \citep{turner_w123}. In each case, the only free parameters in the fit, other than those describing a quadratic model for the out-of-transit systematics, were the transit mid-time and the planet-to-star radius ratio.



\section{Results}

\autoref{fig:RA/DEC} shows the position in the sky of the planets we observed and the coverage of the Ariel mission. Ariel will have continuous viewing zones at the ecliptic poles and, while none of the planets lie within it, HAT-P-18 b is the closest, meaning there will be many potential observing windows for this planet. While Ariel and JWST will be able to observe the whole sky, Twinkle's field of regard is limited to planets within $\pm$40$^\circ$ of the ecliptic plane, meaning HAT-P-18\,b, HAT-P-55\,b and WASP-61\,b cannot be studied by this mission.

\begin{figure*}
    \centering
    \includegraphics[width=0.31\textwidth]{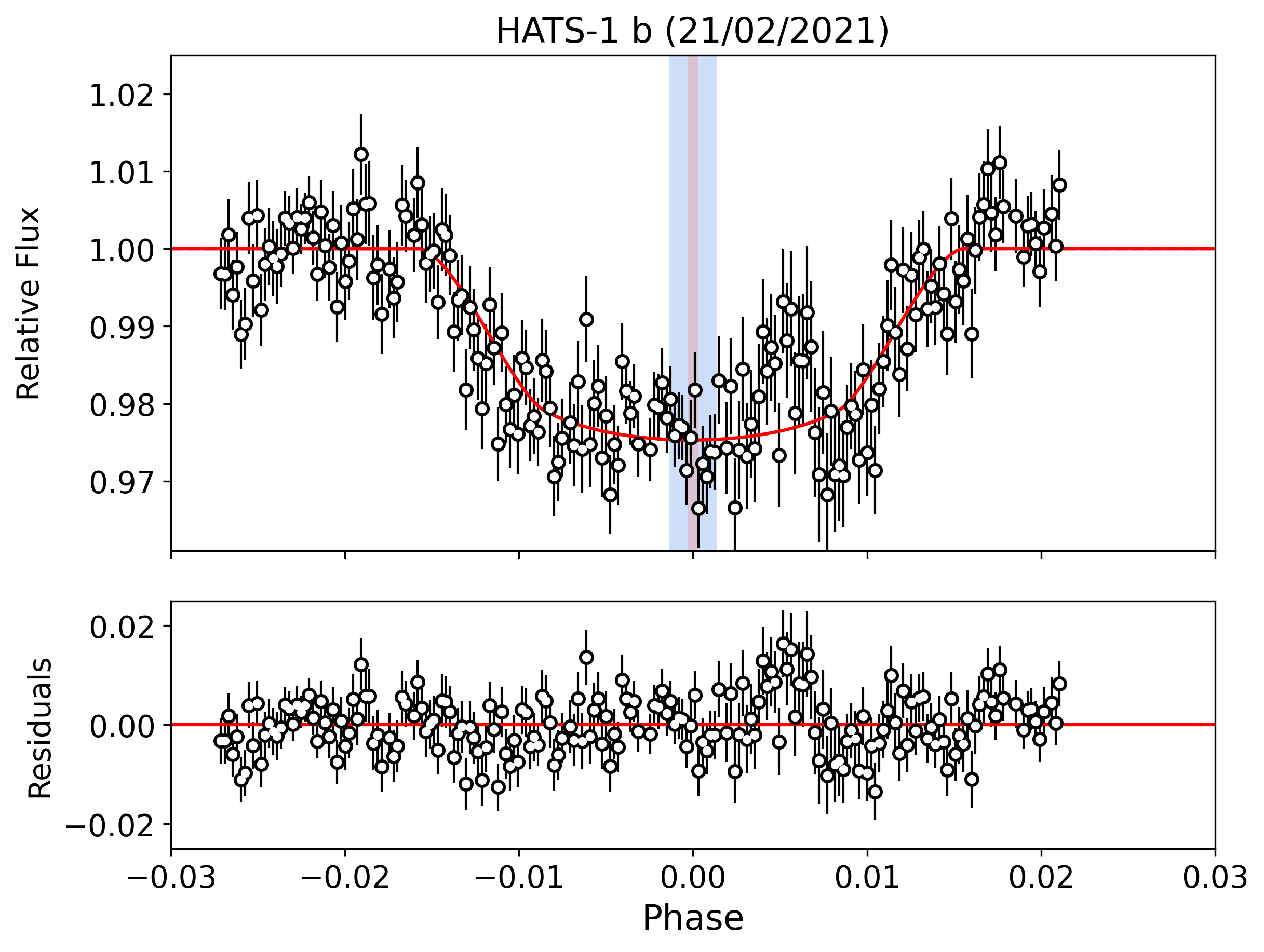}
    \includegraphics[width=0.31\textwidth]{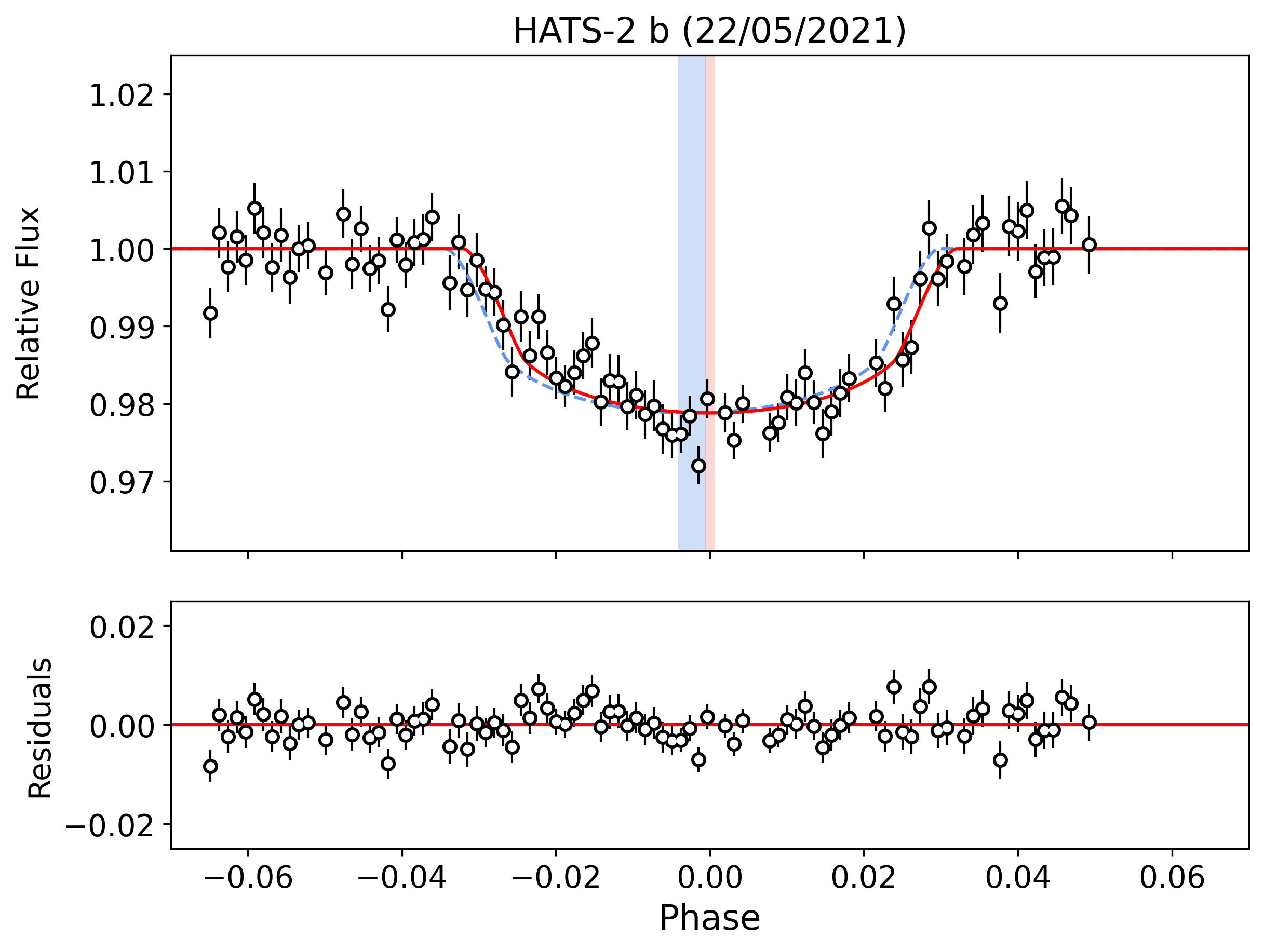}
    \includegraphics[width=0.31\textwidth]{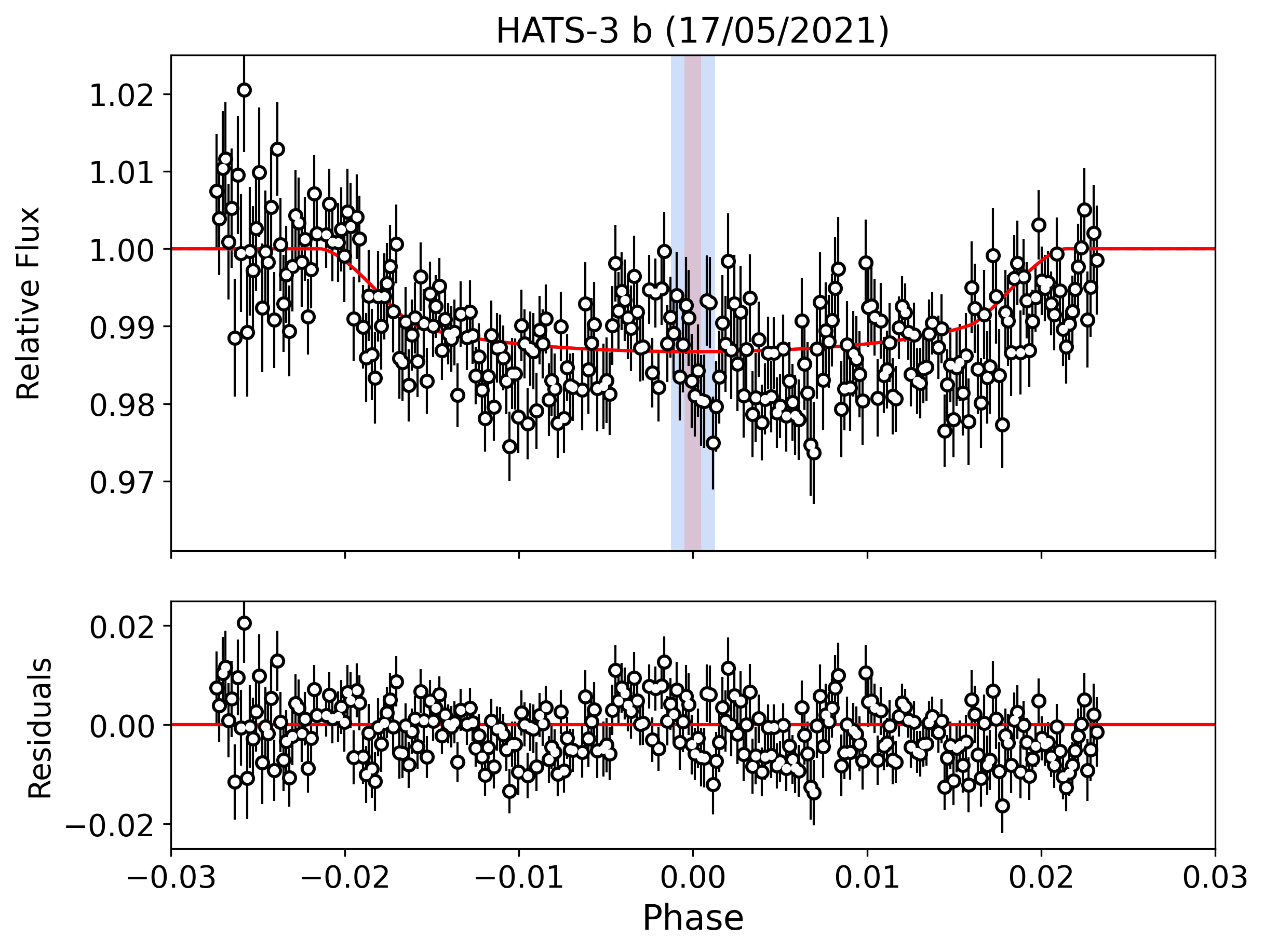}
    \includegraphics[width=0.31\textwidth]{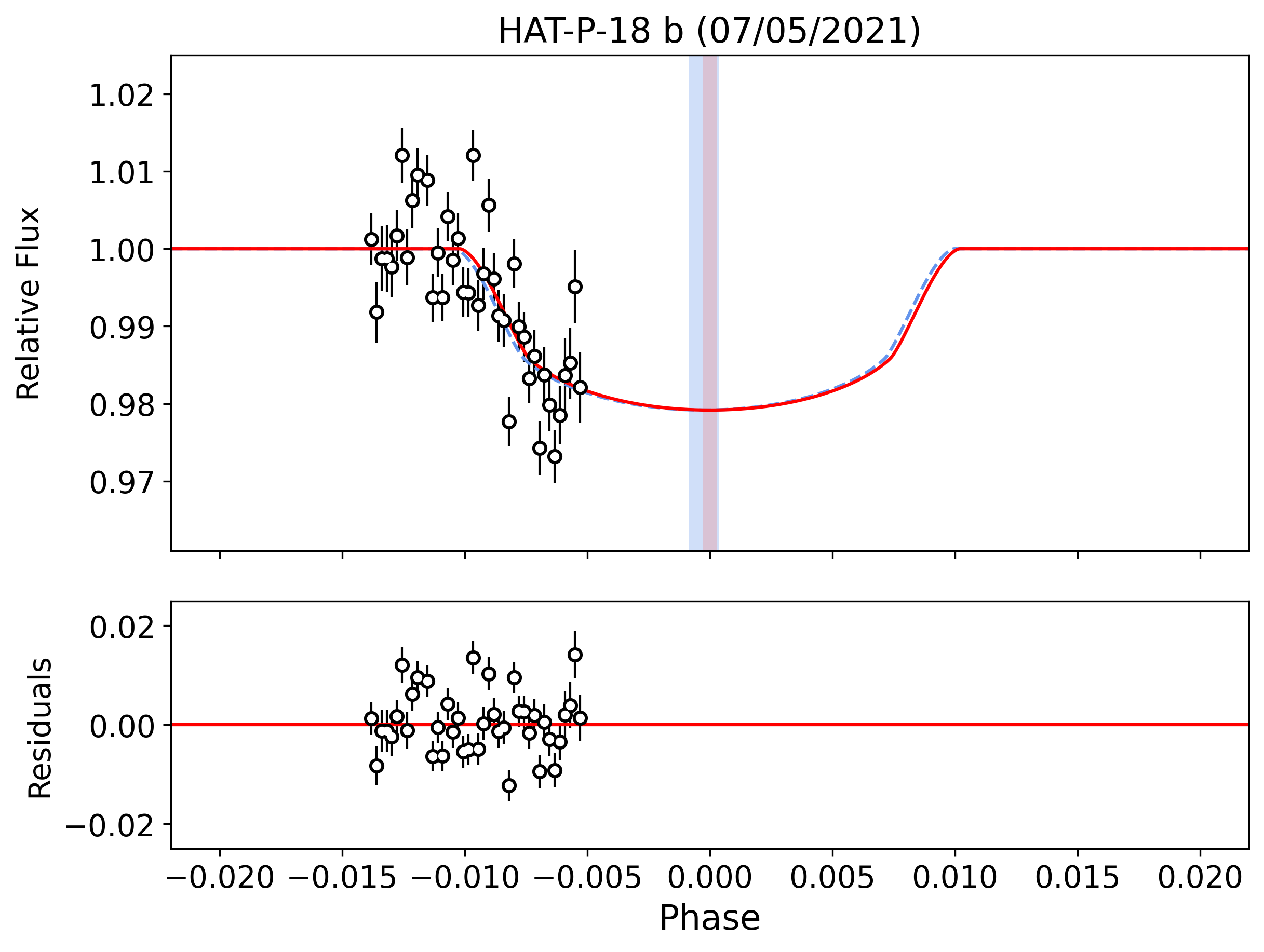}
    \includegraphics[width=0.31\textwidth]{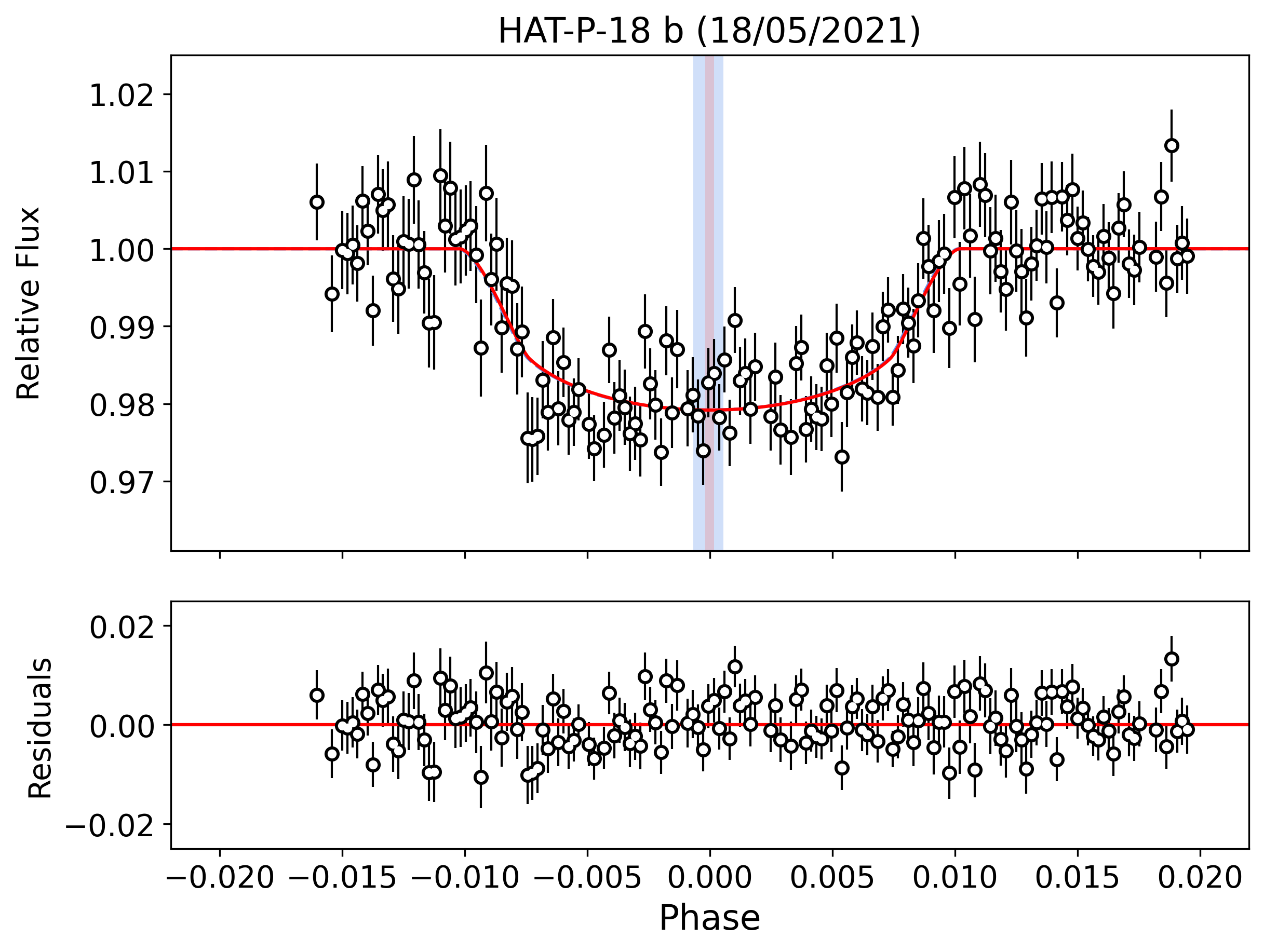}
    \includegraphics[width=0.31\textwidth]{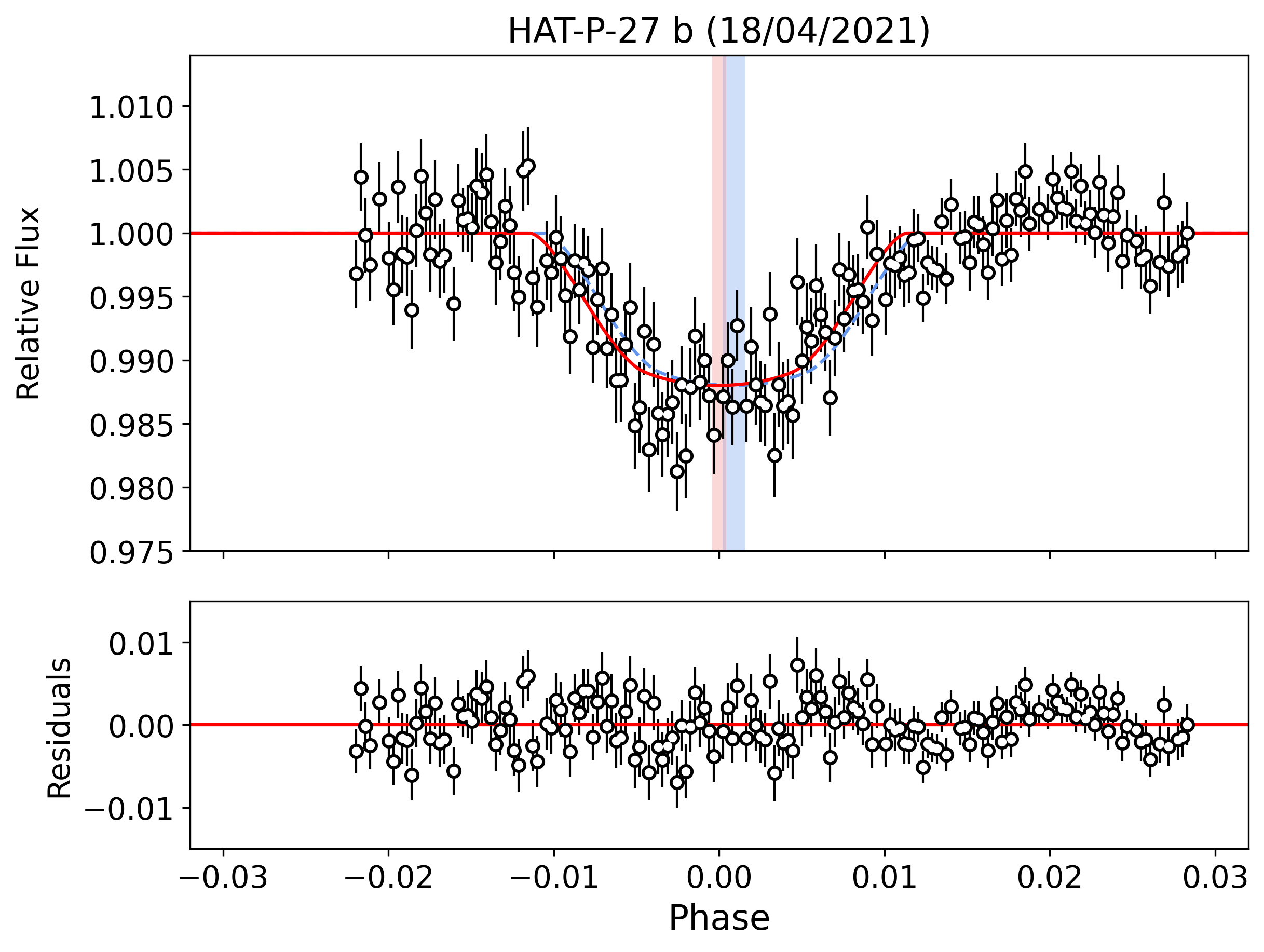}
    \includegraphics[width=0.31\textwidth]{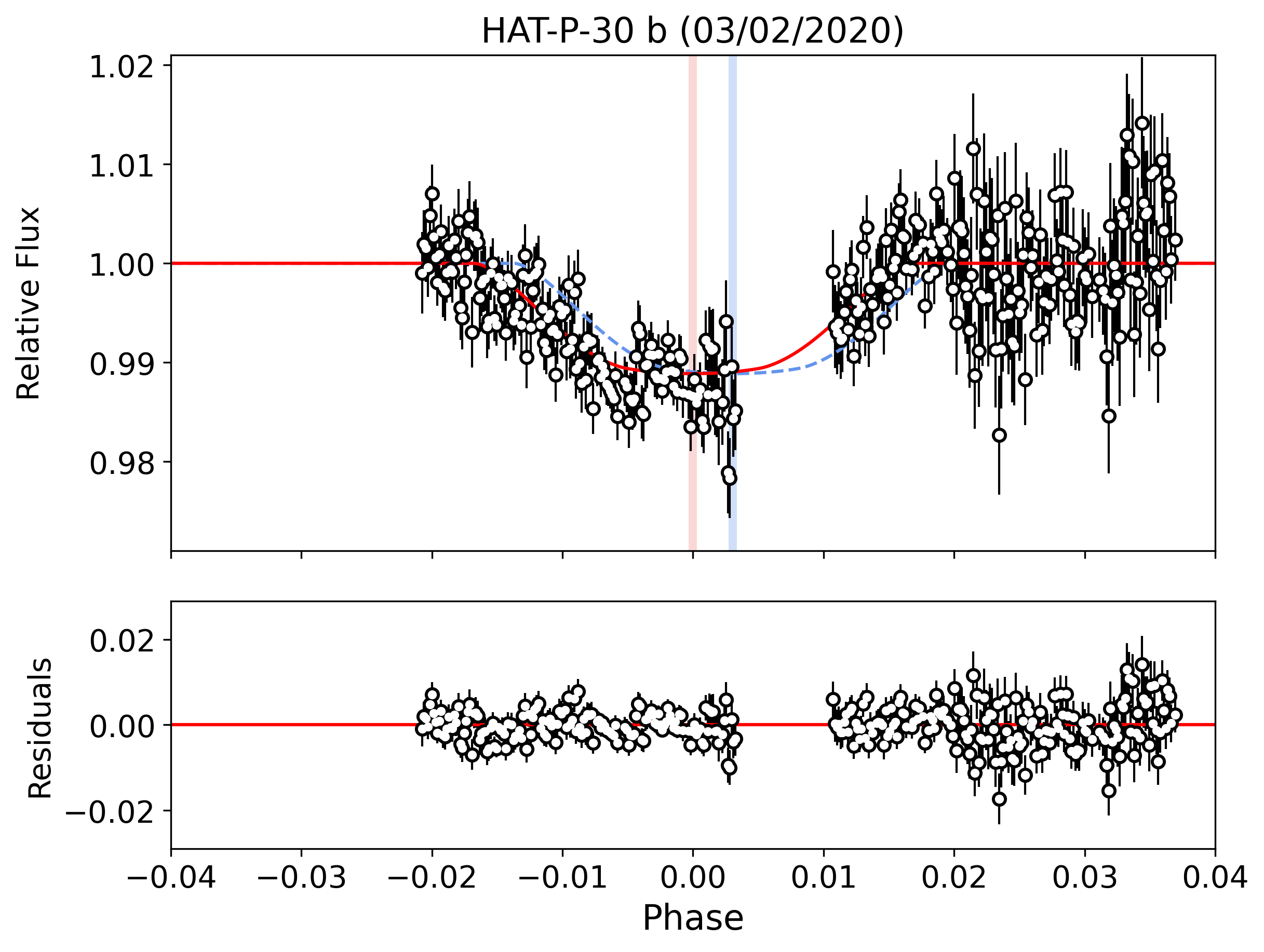}
    \includegraphics[width=0.31\textwidth]{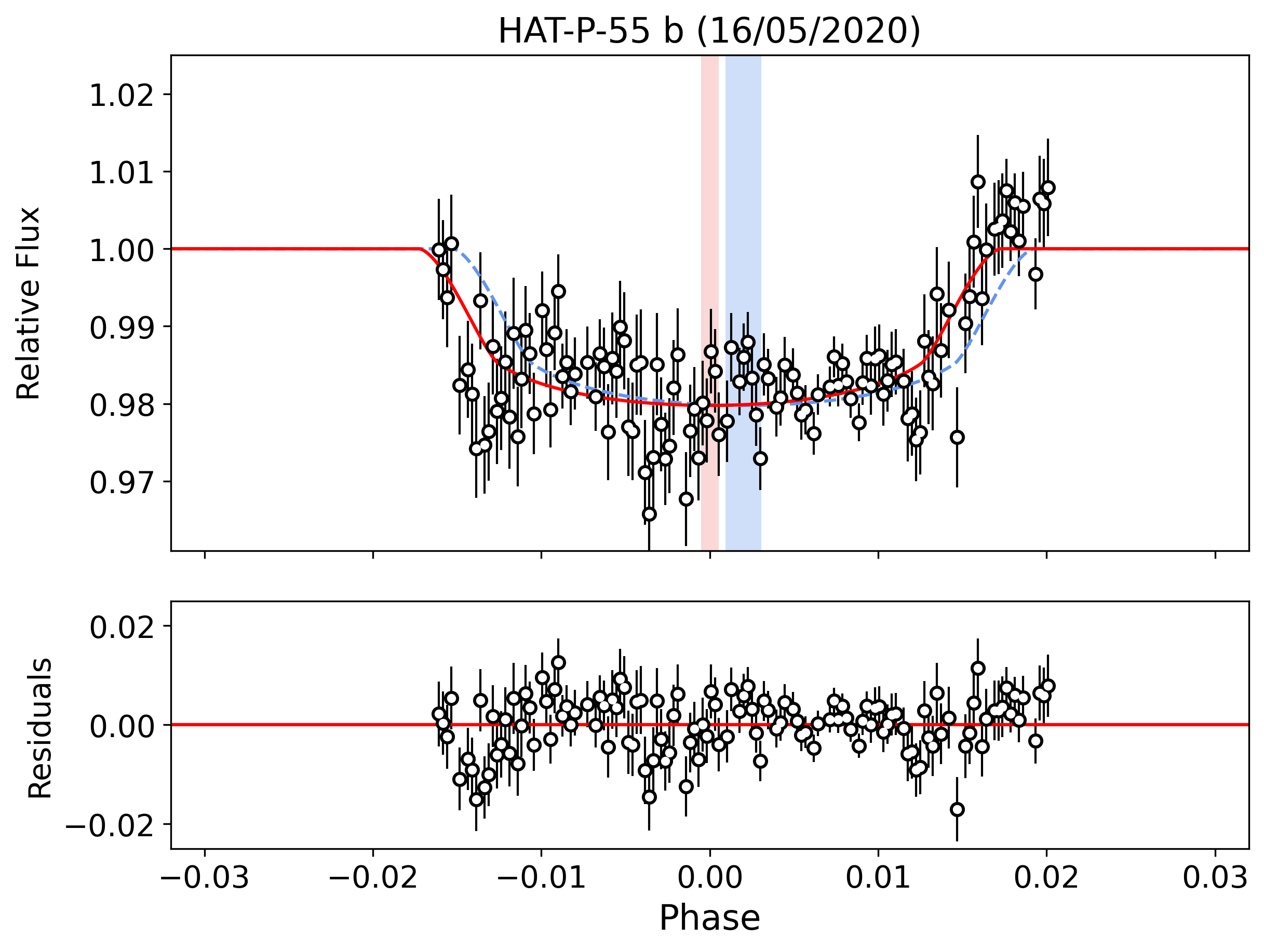}
    \includegraphics[width=0.31\textwidth]{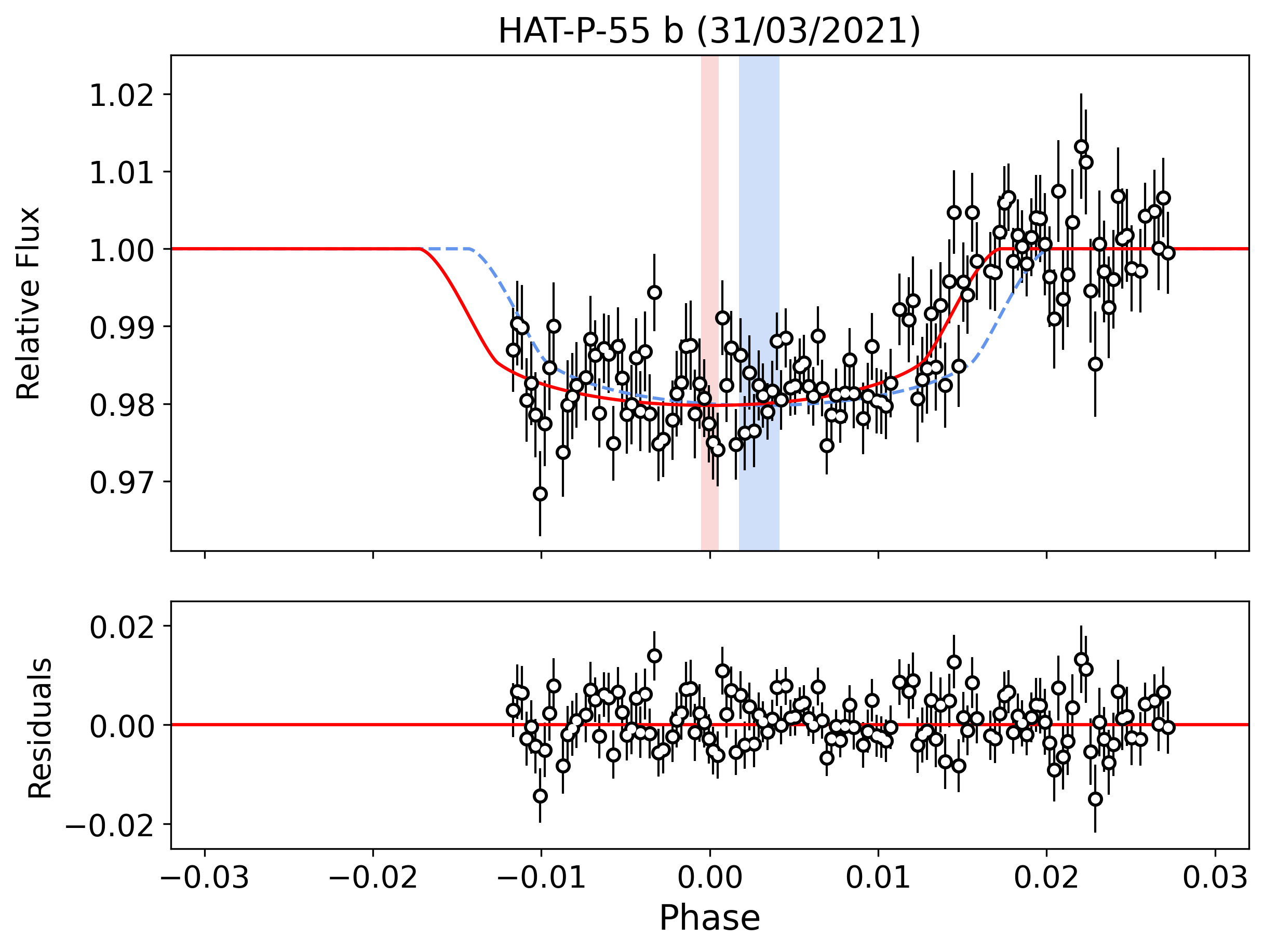}
    \includegraphics[width=0.31\textwidth]{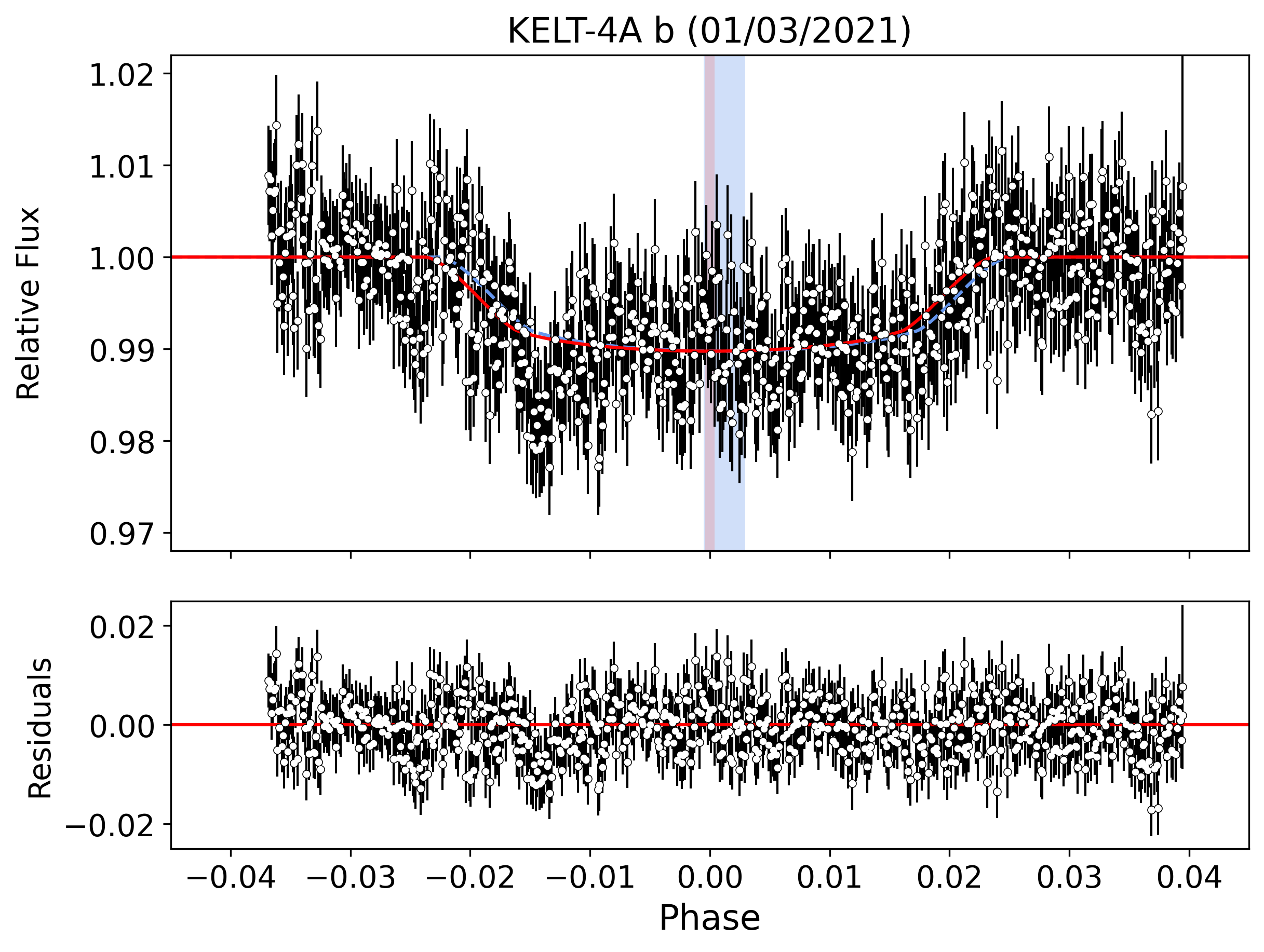}
    \includegraphics[width=0.31\textwidth]{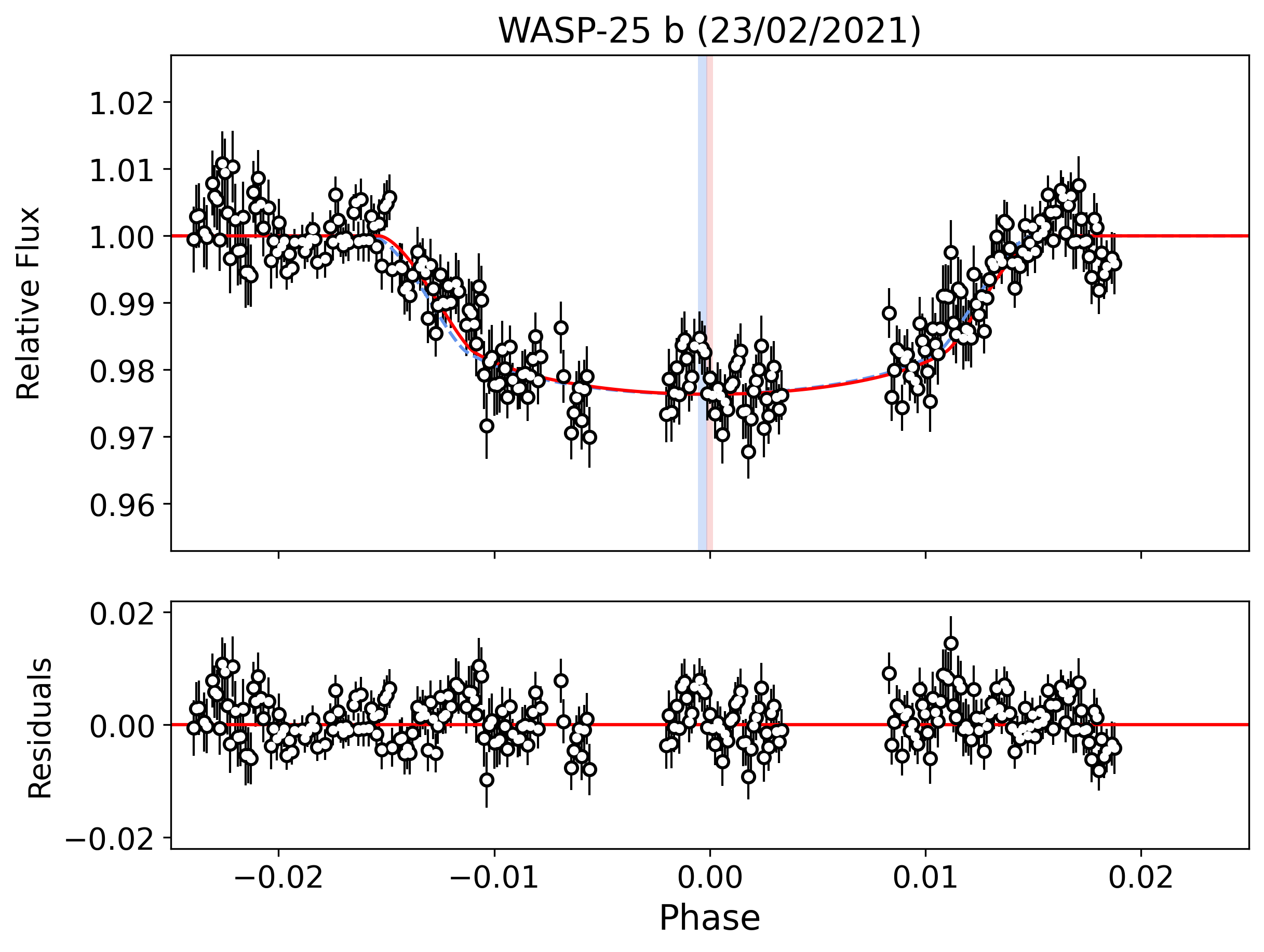}
    \includegraphics[width=0.31\textwidth]{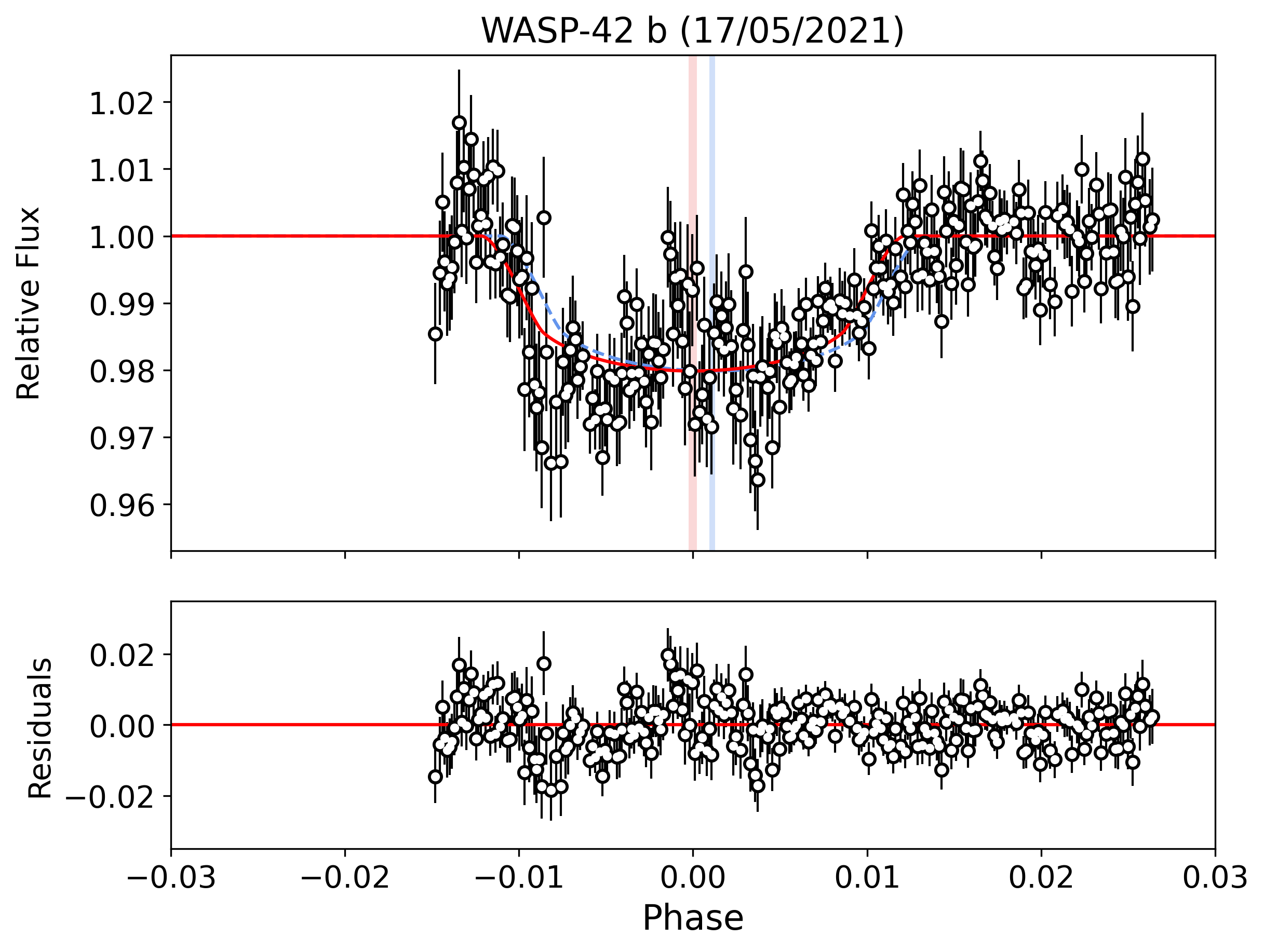}
    \includegraphics[width=0.31\textwidth]{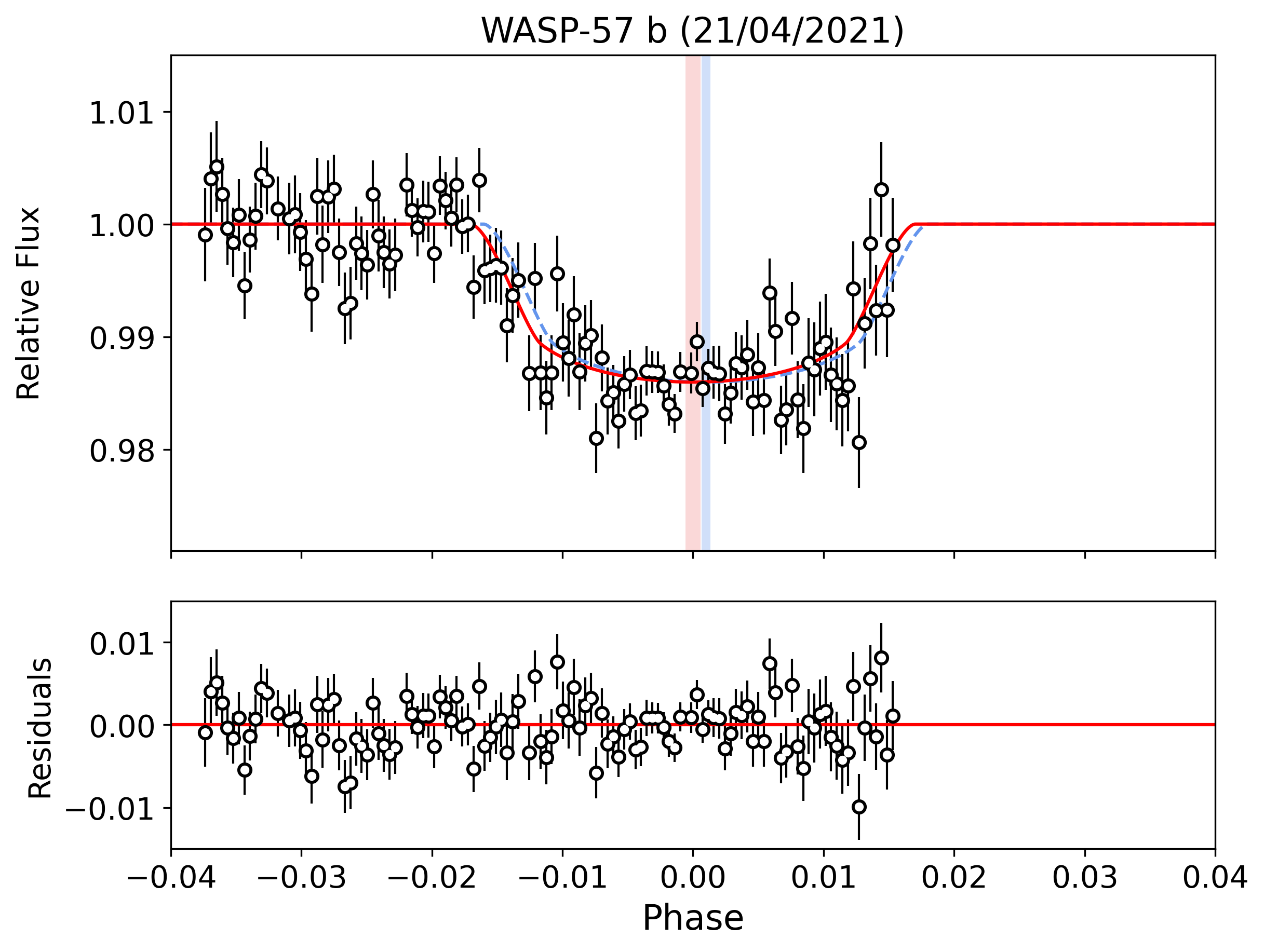}
    \includegraphics[width=0.31\textwidth]{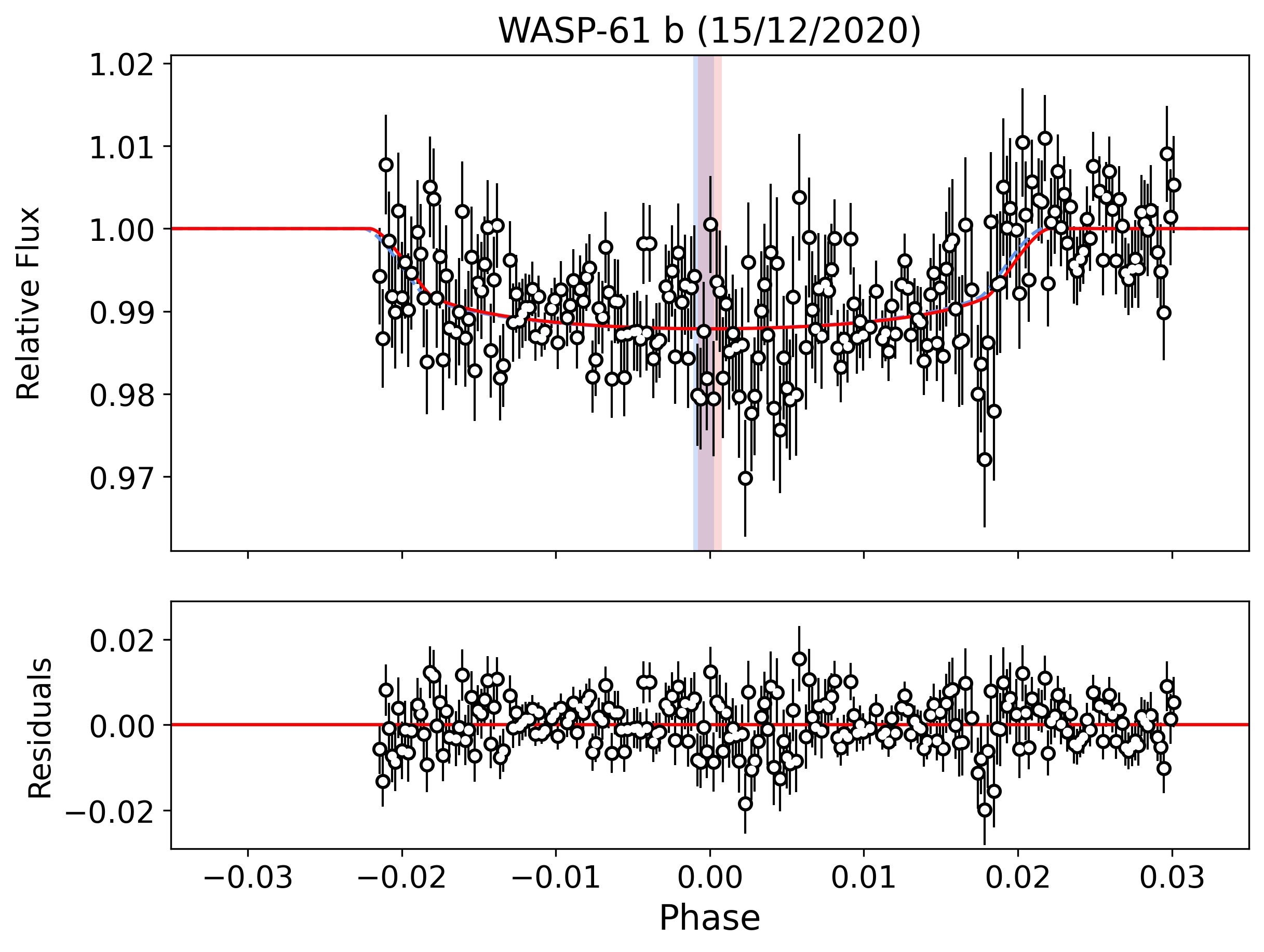}
    \includegraphics[width=0.31\textwidth]{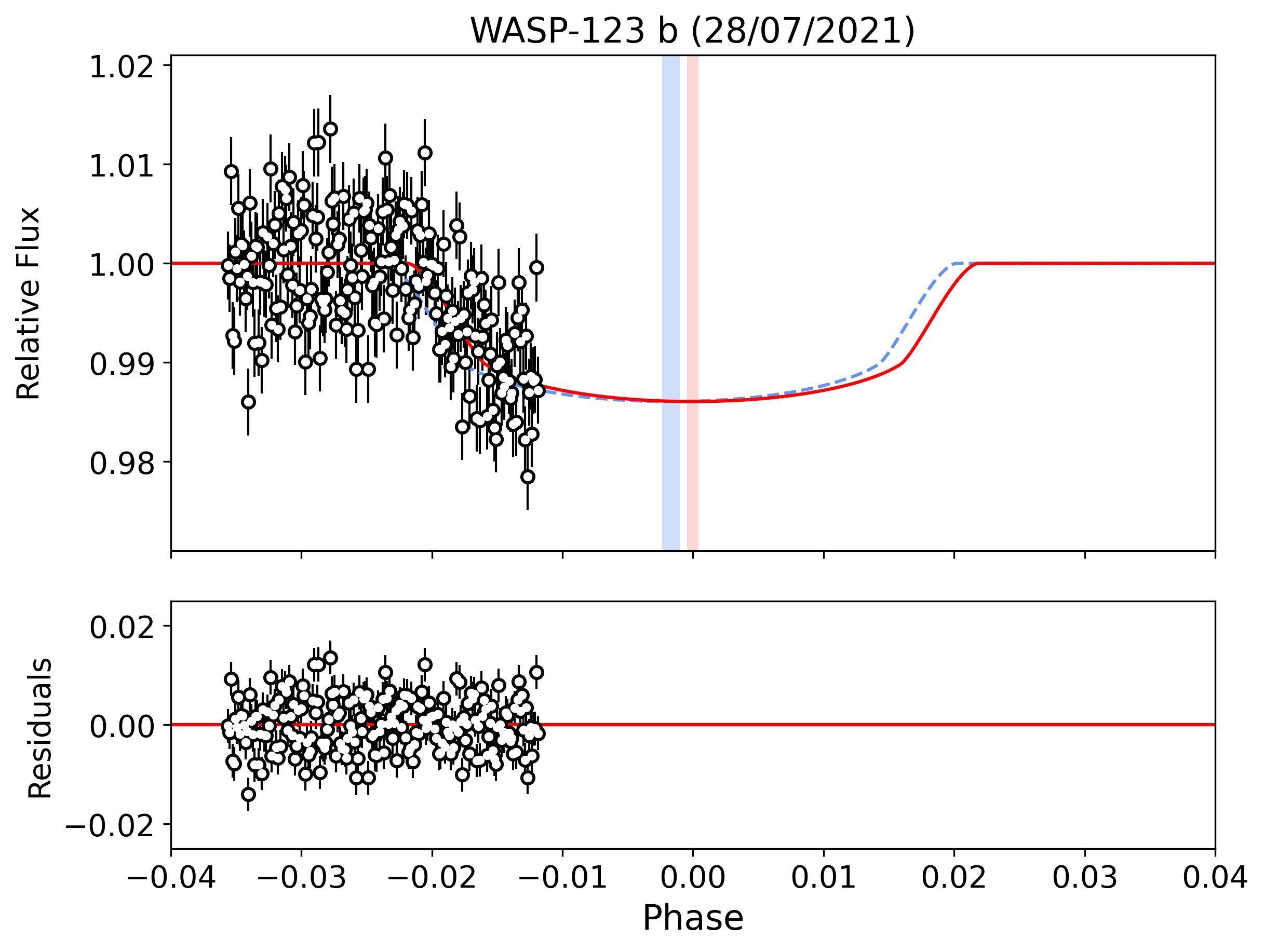}
    \caption{Transit light curves obtained during this project. In each case, the data is shown in
    black with the best-fit transit model in red and the predicted transit model represented by a blue dashed line. For each observation, the red filled region indicates the fitted mid-time and associated uncertainty. Meanwhile, the blue filled region represents the predicted transit mid-time and current uncertainty on this.} 
    \label{fig:results}
\end{figure*}

Across these thirteen planets, our project acquired fifteen transit light curves and the final fits of these are given in \autoref{fig:results}. In each case, the best-fit transit model is given in red while the shaded regions indicate the time window of the fitted mid-time (red) and expected mid-time (blue). Additionally, the expected transit light curve is indicated by a dashed blue line. For each observation, we compared the fitted mid-time to the expected, calculating the observed minus calculated residual (O-C). The transit mid-times and O-C values are given in \autoref{tab:results}.

We note that, for the first observation of HAT-P-18\,b and our observation of WASP-123\,b, HOPS struggled to fit the data when the planet-to-star radius ratio was a free parameter due to the poor coverage of the transit. Therefore, we attempted fitting the transit with a fixed planet-to-star ratio. However, due to the reasons discussed below, we do not report the mid-times in \autoref{tab:results} though the light curve fits are shown in \autoref{fig:results} for completeness.


\section{Discussion}


The quality of our light curves varied between targets but in many cases the transit can be clearly seen and is well-fitted with no significant correlations within the residuals. The precision achieved on the transit mid-time varies from less than a minute to nearly 3 minutes. We have identified a number of potential reasons for this, as discussed below.

\begin{table}[]
    \centering
    \caption{Transit mid-times for each light curve analysed in this project as well as the subsequent observed minus calculated (O-C) residual.}
    \begin{tabular}{ccc}\hline \hline
        Planet &  Mid-time [BJD$_{\rm TDB}$] & O-C [min] \\\hline 
         HATS-1\,b & 2459266.7521$^{+0.0008}_{-0.0010}$ & -0.0$^{+1.2}_{-1.4}$ \\
    HATS-2\,b & 2459268.1531$^{+0.0008}_{-0.0007}$ & 4.4$^{+1.2}_{-1.1}$ \\ 
    HATS-3\,b & 2459352.5818$^{+0.0017}_{-0.0015}$ & -0.1$^{+2.4}_{-2.1}$\\ 
    HAT-P-18\,b & 2459352.7832$^{+0.0010}_{-0.0009}$ & 0.2$^{+1.5}_{-1.3}$ \\ 
    HAT-P-27\,b & 2459322.8860$^{+0.0013}_{-0.0009}$ & -3.9$^{+1.8}_{-1.2}$\\ 
    HAT-P-30\,b & 2458882.5893$^{+0.0009}_{-0.0009}$ & -12.3$^{+1.3}_{-1.3}$\\ 
    HAT-P-55\,b & 2458985.9487$^{+0.0015}_{-0.0015}$ & -10$^{+2.1}_{-1.9}$\\ 
    HAT-P-55\,b & 2459305.0323$^{+0.0016}_{-0.0019}$ & -14.8$^{+2.4}_{-2.7}$ \\ 
    KELT-4A\,b & 2459275.5584$^{+0.0012}_{-0.0011}$ & -5.8$^{+1.8}_{-1.6}$\\ 
    WASP-25\,b & 2459269.4859$^{+0.0005}_{-0.0006}$ & 1.9$^{+0.7}_{-0.9}$\\
    WASP-42\,b  & 2459351.9514$^{+0.0012}_{-0.0011}$ & -8.0$^{+1.8}_{-1.6}$\\ 
    WASP-57\,b  & 2459326.1416$^{+0.0016}_{-0.0014}$ & -3.9$^{+2.3}_{-2.0}$\\
    WASP-61\,b  & 2459198.7400$^{+0.0030}_{-0.0030}$ & 2.8$^{+3.7}_{-4.3}$\\
    \hline\hline 
    \end{tabular}
    \label{tab:results}
\end{table}

Firstly, while many of our light curves cover the full transit duration, plus a baseline of up to toughly half the transit duration, several were interrupted due to bad weather, reducing the coverage. For instance, our observations of HATS-3\,b and WASP-57\,b do not have many data points post-egress while both observations of HAT-P-55\,b do not have any pre-ingress data. However, the worst affected amongst our light curves were the observations of HAT-P-18\,b on 7$^{\rm{th}}$ May 2021 and WASP-123\,b on 28$^{\rm{th}}$ July 2021 as these datasets only covered the ingress and a small amount of time pre-ingress. Despite this, the precision on the transit mid-time in each case is not the worst amongst the sample ($\pm$2.1 and $\pm$2.0 minutes respectively). Nevertheless, as these fits had one fewer free parameter this could have led to an under-estimation of the uncertainty on the transit mid time.

A second cause of differences in the precision of our observations is the brightness of the host star which dictates the amount of flux received. While a brighter star leads to additional flux, if it is too bright the detector must be read-out at a faster rate, leading to a lower duty cycle as the detector spends more time being read. Additionally, very short exposures could be dominated by read noise instead of photon noise from the star. Each of these would lead to a poorer overall quality for the light curve with HATS-3\,b, HAT-P-30\,b, KELT-4A\,b, WASP-42\,b and WASP-123\,b, the five planets with the brightest host stars, being good examples of this.

Additionally, the depth of the planet's transit will affect the ability of HOPS to accurately fit the mid-time. A deeper transit leads to a higher signal-to-noise ratio, allowing the start and end of the transit to be more easily discerned. Examples are HAT-P-27\,b and WASP-25\,b, where both hosts stars have a similar magnitude (11.98 mag and 11.82 respectively), but the transit depths differ (13 mmag and 20 mmag respectively), hence the mid-time of WASP-25\,b has a higher precision than that of HAT-P-27\,b. However, here the transit duration may also be having an effect as HAT-P-27\,b also has a shorter transit time (1.68 hours) than WASP-25\,b (2.76 hours). Also, as HAT-P-18\,b has a relatively deep transit, this could explain the relatively higher precision on the mid-time obtained from the visit of the 7th May although constraints on the mid-time of partial transits are often over-estimated.

Finally, other effects which are harder to quantify are likely to be affecting our observations. One such effect is the airmass: the amount of atmosphere we were observing the star through, and particularly changes in this over the observation period. Furthermore, the phase of the moon and proximity to the host star could also have an effect by increasing the background noise. The comparison star chosen will also affect the precision of our light curves and, while some host stars had many potential comparison stars to chose from, others did not. In truth, the comparative precision of our mid-times is a combination of all these effects, with the dominate cause of a higher uncertainty being hard to determine definitively.

The majority of the planets had observed mid-times which were within 1$\sigma$ of the expected transit time. However, for several planets we found significant offsets and of particular note are HAT-P-30\,b, HAT-P-55\,b and WASP-42\,b. Our observed O-C residuals for HAT-P-30\,b and HAT-P-55\,b are consistent with those found by other ExoClock users\footnote{\cite{kokori_exoclock} and \url{https://www.exoclock.space/database/planets/HAT-P-55b/}}, giving us confidence that the deviation from the literature ephemeris that we find is correct. The cause of this deviation could be a slight inaccuracy on the linear period derived in other works or an actual variation from the linear period due to effects such as orbital decay or transit timing variations from other, as yet unidentified, planets in the system.

On the other hand, WASP-42\,b does not have, at the time of writing, any observations listed on ExoClock. Therefore, further transits will need to be observed to verify our results and WASP-123\,b provides a good example of the need of multiple transits to confirm an O-C drift. Our observation of WASP-123\,b suggested an O-C of 5.8$\pm$2.0 minutes and, while we already questioned the reliability of this result due to the poor transit coverage, this was compounded by recent observations uploaded to ExoClock by other observations which found an O-C of 2.8$\pm$0.5 minutes. Due to this, we choose not to report our mid-time for WASP-123\,b. Likewise, our first observation of HAT-P-18\,b, which had poor coverage, was over 2 minutes different from that found by the ExoClock observations to date and therefore the mid-time is not reported here.

As, for the majority of the planets, we only observed a single transit we do not update the period as part of this work and simply provide the mid-times so that they can be utilised in future studies, such as those that are being regularly produced via the ExoClock programme \citep{kokori_exoclock}.

All the planets observed here are excellent targets for atmospheric characterisation with upcoming facilities. In fact, HAT-P-18\,b has already been studied through transit spectroscopy using the Hubble Space Telescope \citep{tsiaras_30planets}, the William Herschel Telescope \citep{kirk_h18}, and the Hale Telescope at Palomar Observatory \citep{paragas_h18}. These, respectively, have identified the presence of water vapour, observed the effects of Rayleigh scattering, and detected helium in the atmosphere of HAT-P-18\,b. Our work, combined with that of previous ORBYTS projects \citep{edwards_orbytsII,edwards_orbytsI} and other ephemeris follow-up projects \citep[e.g.][]{poddany_etd,mallonn_ephm,zellem_exowatch,kokori_exoclock} increases confidence in the knowledge that these planets will transit at the expected time, hence aiding the study of their atmospheres.

\section{Conclusion}

We present observations of thirteen exoplanets which were rated by the ExoClock site as medium or high priority for photometric follow-up. All these planets are potential targets for future space-based facilities and our observations will help ensure their transit ephemerides are well-known. As TESS and other surveys continue to find planets, ephemeris refinement projects will become ever more important and educational outreach and citizen-science programmes have the potential to play a large role in maintaining transit times for the next generation of telescopes.

\section*{Funding}


The project was undertaken as part of the ORBYTS programme, a public engagement project which partners scientists with schools to support students' involvement in space research. Billy Edwards and Giovanna Tinetti acknowledge funding from the STFC grants ST/S002634/1 and ST/W00254X/1. Cynthia Ho and Hannah Osborne thank the STFC for support through PhD studentships.

\section*{Acknowledgements}


Billy Edwards is the PI of the LCOGT Global Sky Partners 2020 project ``Refining Exoplanet Ephemerides" and thanks the LCOGT network and its coordinators for providing telescope access. Without this access and support, the project presented here would not have been possible. We utilised the 0.4\,m telescopes at Cerro Tololo Observatory, Haleakala Observatory, McDonald Observatory, Sutherland Observatory, Sliding Spring Observatory, and Teide Observatory, each time using their imaging mode and the SDSS-rp filter. The observation of HAT-P-55\,b from 2020 was taken via the Faulkes Telescope Project which is coordinated by Cardiff University and Swansea University. 

The authors wish to thank Matt Densham, from the London Academy of Excellence, and Jon Barker, from Highams Park School, for their dedication in organising the outreach sessions, devoting their spare time for the benefit of their students.

\section*{Software}

Astropy \citep{astropy}, cartopy \citep{cartopy}, corner \citep{corner}, emcee \citep{emcee}, ExoTETHyS \citep{morello_exotethys}, HOPS \citep{tsiaras_hops}, Matplotlib \citep{Hunter_matplotlib}, Numpy \citep{oliphant_numpy}, Pandas \citep{mckinney_pandas}, pylightcurve \citep{tsiaras_plc},  SciPy \citep{scipy}.





\bibliography{main}{}

\begin{thebibliography}{}

\bibitem [\protect \citeauthoryear {%
{Anderson}%
\ \protect \BOthers {.}}{%
{Anderson}%
\ \protect \BOthers {.}}{%
{\protect \APACyear {2011}}%
}]{%
anderson_h27}
\APACinsertmetastar {%
anderson_h27}%
\begin{APACrefauthors}%
{Anderson}, D\BPBI R.%
, {Barros}, S\BPBI C\BPBI C.%
, {Boisse}, I.%
, {Bouchy}, F.%
, {Collier Cameron}, A.%
, {Faedi}, F.%
\BDBL {}{Skillen}, I.%
\end{APACrefauthors}%
\unskip\
\newblock
\APACrefYearMonthDay{2011}{{\APACmonth{05}}}{}.
\newblock
{\BBOQ}\APACrefatitle {{WASP-40b: Independent Discovery of the 0.6 M$_{Jup}$
  Transiting Exoplanet HAT-P-27b}} {{WASP-40b: Independent Discovery of the 0.6
  M$_{Jup}$ Transiting Exoplanet HAT-P-27b}}.{\BBCQ}
\newblock
\APACjournalVolNumPages{\pasp}{123}{903}{555}.
\newblock
\begin{APACrefDOI} \doi{10.1086/660135} \end{APACrefDOI}
\PrintBackRefs{\CurrentBib}

\bibitem [\protect \citeauthoryear {%
{Astropy Collaboration}%
\ \protect \BOthers {.}}{%
{Astropy Collaboration}%
\ \protect \BOthers {.}}{%
{\protect \APACyear {2018}}%
}]{%
astropy}
\APACinsertmetastar {%
astropy}%
\begin{APACrefauthors}%
{Astropy Collaboration}%
, {Price-Whelan}, A\BPBI M.%
, {Sip{\H{o}}cz}, B\BPBI M.%
, {G{\"u}nther}, H\BPBI M.%
, {Lim}, P\BPBI L.%
, {Crawford}, S\BPBI M.%
\BDBL {}{Astropy Contributors}%
\end{APACrefauthors}%
\unskip\
\newblock
\APACrefYearMonthDay{2018}{{\APACmonth{09}}}{}.
\newblock
{\BBOQ}\APACrefatitle {{The Astropy Project: Building an Open-science Project
  and Status of the v2.0 Core Package}} {{The Astropy Project: Building an
  Open-science Project and Status of the v2.0 Core Package}}.{\BBCQ}
\newblock
\APACjournalVolNumPages{\aj}{156}{3}{123}.
\newblock
\begin{APACrefDOI} \doi{10.3847/1538-3881/aabc4f} \end{APACrefDOI}
\PrintBackRefs{\CurrentBib}

\bibitem [\protect \citeauthoryear {%
{Bayliss}%
\ \protect \BOthers {.}}{%
{Bayliss}%
\ \protect \BOthers {.}}{%
{\protect \APACyear {2013}}%
}]{%
bayliss_hats3}
\APACinsertmetastar {%
bayliss_hats3}%
\begin{APACrefauthors}%
{Bayliss}, D.%
, {Zhou}, G.%
, {Penev}, K.%
, {Bakos}, G\BPBI {\'A}.%
, {Hartman}, J\BPBI D.%
, {Jord{\'a}n}, A.%
\BDBL {}{S{\'a}ri}, P.%
\end{APACrefauthors}%
\unskip\
\newblock
\APACrefYearMonthDay{2013}{{\APACmonth{11}}}{}.
\newblock
{\BBOQ}\APACrefatitle {{HATS-3b: An Inflated Hot Jupiter Transiting an F-type
  Star}} {{HATS-3b: An Inflated Hot Jupiter Transiting an F-type Star}}.{\BBCQ}
\newblock
\APACjournalVolNumPages{\aj}{146}{5}{113}.
\newblock
\begin{APACrefDOI} \doi{10.1088/0004-6256/146/5/113} \end{APACrefDOI}
\PrintBackRefs{\CurrentBib}

\bibitem [\protect \citeauthoryear {%
{Bean}%
\ \protect \BOthers {.}}{%
{Bean}%
\ \protect \BOthers {.}}{%
{\protect \APACyear {2018}}%
}]{%
bean_ers}
\APACinsertmetastar {%
bean_ers}%
\begin{APACrefauthors}%
{Bean}, J\BPBI L.%
, {Stevenson}, K\BPBI B.%
, {Batalha}, N\BPBI M.%
, {Berta-Thompson}, Z.%
, {Kreidberg}, L.%
, {Crouzet}, N.%
\BDBL {}{Zingales}, T.%
\end{APACrefauthors}%
\unskip\
\newblock
\APACrefYearMonthDay{2018}{{\APACmonth{11}}}{}.
\newblock
{\BBOQ}\APACrefatitle {{The Transiting Exoplanet Community Early Release
  Science Program for JWST}} {{The Transiting Exoplanet Community Early Release
  Science Program for JWST}}.{\BBCQ}
\newblock
\APACjournalVolNumPages{\pasp}{130}{993}{114402}.
\newblock
\begin{APACrefDOI} \doi{10.1088/1538-3873/aadbf3} \end{APACrefDOI}
\PrintBackRefs{\CurrentBib}

\bibitem [\protect \citeauthoryear {%
{B{\'e}ky}%
\ \protect \BOthers {.}}{%
{B{\'e}ky}%
\ \protect \BOthers {.}}{%
{\protect \APACyear {2011}}%
}]{%
beky_h27}
\APACinsertmetastar {%
beky_h27}%
\begin{APACrefauthors}%
{B{\'e}ky}, B.%
, {Bakos}, G\BPBI {\'A}.%
, {Hartman}, J.%
, {Torres}, G.%
, {Latham}, D\BPBI W.%
, {Jord{\'a}n}, A.%
\BDBL {}{S{\'a}ri}, P.%
\end{APACrefauthors}%
\unskip\
\newblock
\APACrefYearMonthDay{2011}{{\APACmonth{06}}}{}.
\newblock
{\BBOQ}\APACrefatitle {{HAT-P-27b: A Hot Jupiter Transiting a G Star on a 3 Day
  Orbit}} {{HAT-P-27b: A Hot Jupiter Transiting a G Star on a 3 Day
  Orbit}}.{\BBCQ}
\newblock
\APACjournalVolNumPages{\apj}{734}{2}{109}.
\newblock
\begin{APACrefDOI} \doi{10.1088/0004-637X/734/2/109} \end{APACrefDOI}
\PrintBackRefs{\CurrentBib}

\bibitem [\protect \citeauthoryear {%
{Borucki}%
\ \protect \BOthers {.}}{%
{Borucki}%
\ \protect \BOthers {.}}{%
{\protect \APACyear {2010}}%
}]{%
borucki}
\APACinsertmetastar {%
borucki}%
\begin{APACrefauthors}%
{Borucki}, W\BPBI J.%
, {Koch}, D.%
, {Basri}, G.%
, {Batalha}, N.%
, {Brown}, T.%
, {Caldwell}, D.%
\BDBL {}{Prsa}, A.%
\end{APACrefauthors}%
\unskip\
\newblock
\APACrefYearMonthDay{2010}{{\APACmonth{02}}}{}.
\newblock
{\BBOQ}\APACrefatitle {{Kepler Planet-Detection Mission: Introduction and First
  Results}} {{Kepler Planet-Detection Mission: Introduction and First
  Results}}.{\BBCQ}
\newblock
\APACjournalVolNumPages{Science}{327}{5968}{977}.
\newblock
\begin{APACrefDOI} \doi{10.1126/science.1185402} \end{APACrefDOI}
\PrintBackRefs{\CurrentBib}

\bibitem [\protect \citeauthoryear {%
{Boss}%
}{%
{Boss}%
}{%
{\protect \APACyear {2009}}%
}]{%
boss}
\APACinsertmetastar {%
boss}%
\begin{APACrefauthors}%
{Boss}, A.%
\end{APACrefauthors}%
\unskip\
\newblock
\APACrefYear{2009}.
\newblock
\APACrefbtitle {{The crowded universe : the search for living planets}} {{The
  crowded universe : the search for living planets}}.
\PrintBackRefs{\CurrentBib}

\bibitem [\protect \citeauthoryear {%
{Brown}%
\ \protect \BOthers {.}}{%
{Brown}%
\ \protect \BOthers {.}}{%
{\protect \APACyear {2013}}%
}]{%
brown_lco}
\APACinsertmetastar {%
brown_lco}%
\begin{APACrefauthors}%
{Brown}, T\BPBI M.%
, {Baliber}, N.%
, {Bianco}, F\BPBI B.%
, {Bowman}, M.%
, {Burleson}, B.%
, {Conway}, P.%
\BDBL {}{Willis}, M.%
\end{APACrefauthors}%
\unskip\
\newblock
\APACrefYearMonthDay{2013}{{\APACmonth{09}}}{}.
\newblock
{\BBOQ}\APACrefatitle {{Las Cumbres Observatory Global Telescope Network}}
  {{Las Cumbres Observatory Global Telescope Network}}.{\BBCQ}
\newblock
\APACjournalVolNumPages{\pasp}{125}{931}{1031}.
\newblock
\begin{APACrefDOI} \doi{10.1086/673168} \end{APACrefDOI}
\PrintBackRefs{\CurrentBib}

\bibitem [\protect \citeauthoryear {%
{Campbell}%
, {Walker}%
\BCBL {}\ \BBA {} {Yang}%
}{%
{Campbell}%
\ \protect \BOthers {.}}{%
{\protect \APACyear {1988}}%
}]{%
campbell_1988}
\APACinsertmetastar {%
campbell_1988}%
\begin{APACrefauthors}%
{Campbell}, B.%
, {Walker}, G\BPBI A\BPBI H.%
\BCBL {}\ \BBA {} {Yang}, S.%
\end{APACrefauthors}%
\unskip\
\newblock
\APACrefYearMonthDay{1988}{{\APACmonth{08}}}{}.
\newblock
{\BBOQ}\APACrefatitle {{A Search for Substellar Companions to Solar-type
  Stars}} {{A Search for Substellar Companions to Solar-type Stars}}.{\BBCQ}
\newblock
\APACjournalVolNumPages{\apj}{331}{}{902}.
\newblock
\begin{APACrefDOI} \doi{10.1086/166608} \end{APACrefDOI}
\PrintBackRefs{\CurrentBib}

\bibitem [\protect \citeauthoryear {%
Charbonneau%
, Brown%
, Latham%
\BCBL {}\ \BBA {} Mayor%
}{%
Charbonneau%
\ \protect \BOthers {.}}{%
{\protect \APACyear {2000}}%
}]{%
Charbonneau_2000}
\APACinsertmetastar {%
Charbonneau_2000}%
\begin{APACrefauthors}%
Charbonneau, D.%
, Brown, T\BPBI M.%
, Latham, D\BPBI W.%
\BCBL {}\ \BBA {} Mayor, M.%
\end{APACrefauthors}%
\unskip\
\newblock
\APACrefYearMonthDay{2000}{jan}{}.
\newblock
{\BBOQ}\APACrefatitle {Detection of Planetary Transits Across a Sun-like Star}
  {Detection of planetary transits across a sun-like star}.{\BBCQ}
\newblock
\APACjournalVolNumPages{The Astrophysical Journal}{529}{1}{L45--L48}.
\newblock
\begin{APACrefURL} \url{https://doi.org/10.1086/312457} \end{APACrefURL}
\newblock
\begin{APACrefDOI} \doi{10.1086/312457} \end{APACrefDOI}
\PrintBackRefs{\CurrentBib}

\bibitem [\protect \citeauthoryear {%
{Chubb}%
, {Joseph}%
\BCBL {}\ \protect \BOthers {.}}{%
{Chubb}%
, {Joseph}%
\BCBL {}\ \protect \BOthers {.}}{%
{\protect \APACyear {2018}}%
}]{%
chubb_orbytsI}
\APACinsertmetastar {%
chubb_orbytsI}%
\begin{APACrefauthors}%
{Chubb}, K\BPBI L.%
, {Joseph}, M.%
, {Franklin}, J.%
, {Choudhury}, N.%
, {Furtenbacher}, T.%
, {Cs{\'a}sz{\'a}r}, A\BPBI G.%
\BDBL {}{Sousa-Silva}, C.%
\end{APACrefauthors}%
\unskip\
\newblock
\APACrefYearMonthDay{2018}{{\APACmonth{01}}}{}.
\newblock
{\BBOQ}\APACrefatitle {{MARVEL analysis of the measured high-resolution
  rovibrational spectra of C$_{2}$H$_{2}$}} {{MARVEL analysis of the measured
  high-resolution rovibrational spectra of C$_{2}$H$_{2}$}}.{\BBCQ}
\newblock
\APACjournalVolNumPages{\jqsrt}{204}{}{42-55}.
\newblock
\begin{APACrefDOI} \doi{10.1016/j.jqsrt.2017.08.018} \end{APACrefDOI}
\PrintBackRefs{\CurrentBib}

\bibitem [\protect \citeauthoryear {%
{Chubb}%
, {Naumenko}%
\BCBL {}\ \protect \BOthers {.}}{%
{Chubb}%
, {Naumenko}%
\BCBL {}\ \protect \BOthers {.}}{%
{\protect \APACyear {2018}}%
}]{%
chubb_orbytsII}
\APACinsertmetastar {%
chubb_orbytsII}%
\begin{APACrefauthors}%
{Chubb}, K\BPBI L.%
, {Naumenko}, O.%
, {Keely}, S.%
, {Bartolotto}, S.%
, {Macdonald}, S.%
, {Mukhtar}, M.%
\BDBL {}{Tennyson}, J.%
\end{APACrefauthors}%
\unskip\
\newblock
\APACrefYearMonthDay{2018}{{\APACmonth{10}}}{}.
\newblock
{\BBOQ}\APACrefatitle {{MARVEL analysis of the measured high-resolution
  rovibrational spectra of H$_{2}^{32}$S}} {{MARVEL analysis of the measured
  high-resolution rovibrational spectra of H$_{2}^{32}$S}}.{\BBCQ}
\newblock
\APACjournalVolNumPages{\jqsrt}{218}{}{178-186}.
\newblock
\begin{APACrefDOI} \doi{10.1016/j.jqsrt.2018.07.012} \end{APACrefDOI}
\PrintBackRefs{\CurrentBib}

\bibitem [\protect \citeauthoryear {%
{Darby-Lewis}%
\ \protect \BOthers {.}}{%
{Darby-Lewis}%
\ \protect \BOthers {.}}{%
{\protect \APACyear {2019}}%
}]{%
darby_orbyts}
\APACinsertmetastar {%
darby_orbyts}%
\begin{APACrefauthors}%
{Darby-Lewis}, D.%
, {Shah}, H.%
, {Joshi}, D.%
, {Khan}, F.%
, {Kauwo}, M.%
, {Sethi}, N.%
\BDBL {}{Tennyson}, J.%
\end{APACrefauthors}%
\unskip\
\newblock
\APACrefYearMonthDay{2019}{{\APACmonth{08}}}{}.
\newblock
{\BBOQ}\APACrefatitle {{MARVEL analysis of the measured high-resolution spectra
  of {}1$^{4}$ NH}} {{MARVEL analysis of the measured high-resolution spectra
  of {}1$^{4}$ NH}}.{\BBCQ}
\newblock
\APACjournalVolNumPages{Journal of Molecular Spectroscopy}{362}{}{69-76}.
\newblock
\begin{APACrefDOI} \doi{10.1016/j.jms.2019.06.002} \end{APACrefDOI}
\PrintBackRefs{\CurrentBib}

\bibitem [\protect \citeauthoryear {%
{Dragomir}%
\ \protect \BOthers {.}}{%
{Dragomir}%
\ \protect \BOthers {.}}{%
{\protect \APACyear {2020}}%
}]{%
dragomir}
\APACinsertmetastar {%
dragomir}%
\begin{APACrefauthors}%
{Dragomir}, D.%
, {Harris}, M.%
, {Pepper}, J.%
, {Barclay}, T.%
, {Villanueva}, J., Steven%
, {Ricker}, G\BPBI R.%
\BDBL {}{Yahalomi}, D.%
\end{APACrefauthors}%
\unskip\
\newblock
\APACrefYearMonthDay{2020}{{\APACmonth{05}}}{}.
\newblock
{\BBOQ}\APACrefatitle {{Securing the Legacy of TESS through the Care and
  Maintenance of TESS Planet Ephemerides}} {{Securing the Legacy of TESS
  through the Care and Maintenance of TESS Planet Ephemerides}}.{\BBCQ}
\newblock
\APACjournalVolNumPages{\aj}{159}{5}{219}.
\newblock
\begin{APACrefDOI} \doi{10.3847/1538-3881/ab845d} \end{APACrefDOI}
\PrintBackRefs{\CurrentBib}

\bibitem [\protect \citeauthoryear {%
{Eastman}%
\ \protect \BOthers {.}}{%
{Eastman}%
\ \protect \BOthers {.}}{%
{\protect \APACyear {2016}}%
}]{%
eastman_k4}
\APACinsertmetastar {%
eastman_k4}%
\begin{APACrefauthors}%
{Eastman}, J\BPBI D.%
, {Beatty}, T\BPBI G.%
, {Siverd}, R\BPBI J.%
, {Antognini}, J\BPBI M\BPBI O.%
, {Penny}, M\BPBI T.%
, {Gonzales}, E\BPBI J.%
\BDBL {}{Trueblood}, P.%
\end{APACrefauthors}%
\unskip\
\newblock
\APACrefYearMonthDay{2016}{{\APACmonth{02}}}{}.
\newblock
{\BBOQ}\APACrefatitle {{KELT-4Ab: An Inflated Hot Jupiter Transiting the Bright
  (V {\ensuremath{\sim}} 10) Component of a Hierarchical Triple}} {{KELT-4Ab:
  An Inflated Hot Jupiter Transiting the Bright (V {\ensuremath{\sim}} 10)
  Component of a Hierarchical Triple}}.{\BBCQ}
\newblock
\APACjournalVolNumPages{\aj}{151}{2}{45}.
\newblock
\begin{APACrefDOI} \doi{10.3847/0004-6256/151/2/45} \end{APACrefDOI}
\PrintBackRefs{\CurrentBib}

\bibitem [\protect \citeauthoryear {%
{Edwards}%
\ \protect \BOthers {.}}{%
{Edwards}%
\ \protect \BOthers {.}}{%
{\protect \APACyear {2020}}%
}]{%
edwards_orbytsII}
\APACinsertmetastar {%
edwards_orbytsII}%
\begin{APACrefauthors}%
{Edwards}, B.%
, {Anisman}, L.%
, {Changeat}, Q.%
, {Morvan}, M.%
, {Wright}, S.%
, {Yip}, K\BPBI H.%
\BDBL {}{Tennyson}, J.%
\end{APACrefauthors}%
\unskip\
\newblock
\APACrefYearMonthDay{2020}{{\APACmonth{07}}}{}.
\newblock
{\BBOQ}\APACrefatitle {{Original Research by Young Twinkle Students (Orbyts):
  Ephemeris Refinement of Transiting Exoplanets II}} {{Original Research by
  Young Twinkle Students (Orbyts): Ephemeris Refinement of Transiting
  Exoplanets II}}.{\BBCQ}
\newblock
\APACjournalVolNumPages{Research Notes of the American Astronomical
  Society}{4}{7}{109}.
\newblock
\begin{APACrefDOI} \doi{10.3847/2515-5172/aba42b} \end{APACrefDOI}
\PrintBackRefs{\CurrentBib}

\bibitem [\protect \citeauthoryear {%
{Edwards}%
\ \protect \BOthers {.}}{%
{Edwards}%
\ \protect \BOthers {.}}{%
{\protect \APACyear {2021}}%
}]{%
edwards_orbytsI}
\APACinsertmetastar {%
edwards_orbytsI}%
\begin{APACrefauthors}%
{Edwards}, B.%
, {Changeat}, Q.%
, {Yip}, K\BPBI H.%
, {Tsiaras}, A.%
, {Taylor}, J.%
, {Akhtar}, B.%
\BDBL {}{Tennyson}, J.%
\end{APACrefauthors}%
\unskip\
\newblock
\APACrefYearMonthDay{2021}{{\APACmonth{07}}}{}.
\newblock
{\BBOQ}\APACrefatitle {{Original Research by Young Twinkle Students (ORBYTS):
  Ephemeris Refinement of Transiting Exoplanets}} {{Original Research by Young
  Twinkle Students (ORBYTS): Ephemeris Refinement of Transiting
  Exoplanets}}.{\BBCQ}
\newblock
\APACjournalVolNumPages{\mnras}{504}{4}{5671-5684}.
\newblock
\begin{APACrefDOI} \doi{10.1093/mnras/staa1245} \end{APACrefDOI}
\PrintBackRefs{\CurrentBib}

\bibitem [\protect \citeauthoryear {%
{Edwards}%
, {Mugnai}%
, {Tinetti}%
, {Pascale}%
\BCBL {}\ \BBA {} {Sarkar}%
}{%
{Edwards}%
, {Mugnai}%
\BCBL {}\ \protect \BOthers {.}}{%
{\protect \APACyear {2019}}%
}]{%
ariel_targets}
\APACinsertmetastar {%
ariel_targets}%
\begin{APACrefauthors}%
{Edwards}, B.%
, {Mugnai}, L.%
, {Tinetti}, G.%
, {Pascale}, E.%
\BCBL {}\ \BBA {} {Sarkar}, S.%
\end{APACrefauthors}%
\unskip\
\newblock
\APACrefYearMonthDay{2019}{{\APACmonth{06}}}{}.
\newblock
{\BBOQ}\APACrefatitle {{An Updated Study of Potential Targets for Ariel}} {{An
  Updated Study of Potential Targets for Ariel}}.{\BBCQ}
\newblock
\APACjournalVolNumPages{\aj}{157}{6}{242}.
\newblock
\begin{APACrefDOI} \doi{10.3847/1538-3881/ab1cb9} \end{APACrefDOI}
\PrintBackRefs{\CurrentBib}

\bibitem [\protect \citeauthoryear {%
{Edwards}%
, {Rice}%
\BCBL {}\ \protect \BOthers {.}}{%
{Edwards}%
, {Rice}%
\BCBL {}\ \protect \BOthers {.}}{%
{\protect \APACyear {2019}}%
}]{%
twinkle}
\APACinsertmetastar {%
twinkle}%
\begin{APACrefauthors}%
{Edwards}, B.%
, {Rice}, M.%
, {Zingales}, T.%
, {Tessenyi}, M.%
, {Waldmann}, I.%
, {Tinetti}, G.%
\BDBL {}{Sarkar}, S.%
\end{APACrefauthors}%
\unskip\
\newblock
\APACrefYearMonthDay{2019}{{\APACmonth{04}}}{}.
\newblock
{\BBOQ}\APACrefatitle {{Exoplanet spectroscopy and photometry with the Twinkle
  space telescope}} {{Exoplanet spectroscopy and photometry with the Twinkle
  space telescope}}.{\BBCQ}
\newblock
\APACjournalVolNumPages{Experimental Astronomy}{47}{1-2}{29-63}.
\newblock
\begin{APACrefDOI} \doi{10.1007/s10686-018-9611-4} \end{APACrefDOI}
\PrintBackRefs{\CurrentBib}

\bibitem [\protect \citeauthoryear {%
{Enoch}%
\ \protect \BOthers {.}}{%
{Enoch}%
\ \protect \BOthers {.}}{%
{\protect \APACyear {2011}}%
}]{%
enoch_w25}
\APACinsertmetastar {%
enoch_w25}%
\begin{APACrefauthors}%
{Enoch}, B.%
, {Cameron}, A\BPBI C.%
, {Anderson}, D\BPBI R.%
, {Lister}, T\BPBI A.%
, {Hellier}, C.%
, {Maxted}, P\BPBI F\BPBI L.%
\BDBL {}{Udry}, S.%
\end{APACrefauthors}%
\unskip\
\newblock
\APACrefYearMonthDay{2011}{{\APACmonth{01}}}{}.
\newblock
{\BBOQ}\APACrefatitle {{WASP-25b: a 0.6 M$_{J}$ planet in the Southern
  hemisphere}} {{WASP-25b: a 0.6 M$_{J}$ planet in the Southern
  hemisphere}}.{\BBCQ}
\newblock
\APACjournalVolNumPages{\mnras}{410}{3}{1631-1636}.
\newblock
\begin{APACrefDOI} \doi{10.1111/j.1365-2966.2010.17550.x} \end{APACrefDOI}
\PrintBackRefs{\CurrentBib}

\bibitem [\protect \citeauthoryear {%
{Faedi}%
\ \protect \BOthers {.}}{%
{Faedi}%
\ \protect \BOthers {.}}{%
{\protect \APACyear {2013}}%
}]{%
faedi_w57}
\APACinsertmetastar {%
faedi_w57}%
\begin{APACrefauthors}%
{Faedi}, F.%
, {Pollacco}, D.%
, {Barros}, S\BPBI C\BPBI C.%
, {Brown}, D.%
, {Collier Cameron}, A.%
, {Doyle}, A\BPBI P.%
\BDBL {}{Watson}, C.%
\end{APACrefauthors}%
\unskip\
\newblock
\APACrefYearMonthDay{2013}{{\APACmonth{03}}}{}.
\newblock
{\BBOQ}\APACrefatitle {{WASP-54b, WASP-56b, and WASP-57b: Three new sub-Jupiter
  mass planets from SuperWASP}} {{WASP-54b, WASP-56b, and WASP-57b: Three new
  sub-Jupiter mass planets from SuperWASP}}.{\BBCQ}
\newblock
\APACjournalVolNumPages{\aap}{551}{}{A73}.
\newblock
\begin{APACrefDOI} \doi{10.1051/0004-6361/201220520} \end{APACrefDOI}
\PrintBackRefs{\CurrentBib}

\bibitem [\protect \citeauthoryear {%
{Foreman-Mackey}%
, {Hogg}%
, {Lang}%
\BCBL {}\ \BBA {} {Goodman}%
}{%
{Foreman-Mackey}%
\ \protect \BOthers {.}}{%
{\protect \APACyear {2013}}%
}]{%
emcee}
\APACinsertmetastar {%
emcee}%
\begin{APACrefauthors}%
{Foreman-Mackey}, D.%
, {Hogg}, D\BPBI W.%
, {Lang}, D.%
\BCBL {}\ \BBA {} {Goodman}, J.%
\end{APACrefauthors}%
\unskip\
\newblock
\APACrefYearMonthDay{2013}{{\APACmonth{03}}}{}.
\newblock
{\BBOQ}\APACrefatitle {{emcee: The MCMC Hammer}} {{emcee: The MCMC
  Hammer}}.{\BBCQ}
\newblock
\APACjournalVolNumPages{\pasp}{125}{925}{306}.
\newblock
\begin{APACrefDOI} \doi{10.1086/670067} \end{APACrefDOI}
\PrintBackRefs{\CurrentBib}

\bibitem [\protect \citeauthoryear {%
Foreman-Mackey%
}{%
Foreman-Mackey%
}{%
{\protect \APACyear {2016}}%
}]{%
corner}
\APACinsertmetastar {%
corner}%
\begin{APACrefauthors}%
Foreman-Mackey, D.%
\end{APACrefauthors}%
\unskip\
\newblock
\APACrefYearMonthDay{2016}{jun}{}.
\newblock
{\BBOQ}\APACrefatitle {corner.py: Scatterplot matrices in Python} {corner.py:
  Scatterplot matrices in python}.{\BBCQ}
\newblock
\APACjournalVolNumPages{The Journal of Open Source Software}{1}{2}{24}.
\newblock
\begin{APACrefURL} \url{https://doi.org/10.21105/joss.00024} \end{APACrefURL}
\newblock
\begin{APACrefDOI} \doi{10.21105/joss.00024} \end{APACrefDOI}
\PrintBackRefs{\CurrentBib}

\bibitem [\protect \citeauthoryear {%
Francis%
\ \protect \BOthers {.}}{%
Francis%
\ \protect \BOthers {.}}{%
{\protect \APACyear {2020}}%
}]{%
francis_orbyts}
\APACinsertmetastar {%
francis_orbyts}%
\begin{APACrefauthors}%
Francis, A.%
, Brown, J.%
, Cameron, T.%
, Crawford~Clarke, R.%
, Dodd, R.%
, Hurdle, J.%
\BDBL {}Muller, J\BHBI P.%
\end{APACrefauthors}%
\unskip\
\newblock
\APACrefYearMonthDay{2020}{}{}.
\newblock
{\BBOQ}\APACrefatitle {A Multi-Annotator Survey of Sub-km Craters on Mars} {A
  multi-annotator survey of sub-km craters on mars}.{\BBCQ}
\newblock
\APACjournalVolNumPages{Data}{5}{3}{}.
\newblock
\begin{APACrefURL} \url{https://www.mdpi.com/2306-5729/5/3/70} \end{APACrefURL}
\newblock
\begin{APACrefDOI} \doi{10.3390/data5030070} \end{APACrefDOI}
\PrintBackRefs{\CurrentBib}

\bibitem [\protect \citeauthoryear {%
French%
\ \protect \BOthers {.}}{%
French%
\ \protect \BOthers {.}}{%
{\protect \APACyear {2020}}%
}]{%
french_orbyts}
\APACinsertmetastar {%
french_orbyts}%
\begin{APACrefauthors}%
French, R.%
, James, A.%
, Baker, D.%
, Dunn, W.%
, Matthews, S.%
, da Silva~Pestana, B.%
\BDBL {}Trindade, G.%
\end{APACrefauthors}%
\unskip\
\newblock
\APACrefYearMonthDay{2020}{12}{}.
\newblock
{\BBOQ}\APACrefatitle {{Opening pupils' eyes to the Sun}} {{Opening pupils'
  eyes to the Sun}}.{\BBCQ}
\newblock
\APACjournalVolNumPages{Astronomy \& Geophysics}{61}{6}{6.22-6.23}.
\newblock
\begin{APACrefURL} \url{https://doi.org/10.1093/astrogeo/ataa085}
  \end{APACrefURL}
\newblock
\begin{APACrefDOI} \doi{10.1093/astrogeo/ataa085} \end{APACrefDOI}
\PrintBackRefs{\CurrentBib}

\bibitem [\protect \citeauthoryear {%
{Galle}%
}{%
{Galle}%
}{%
{\protect \APACyear {1846}}%
}]{%
galle_neptune}
\APACinsertmetastar {%
galle_neptune}%
\begin{APACrefauthors}%
{Galle}, J\BPBI G.%
\end{APACrefauthors}%
\unskip\
\newblock
\APACrefYearMonthDay{1846}{{\APACmonth{11}}}{}.
\newblock
{\BBOQ}\APACrefatitle {{Account of the discovery of Le Verrier's planet
  Neptune, at Berlin, Sept. 23, 1846}} {{Account of the discovery of Le
  Verrier's planet Neptune, at Berlin, Sept. 23, 1846}}.{\BBCQ}
\newblock
\APACjournalVolNumPages{\mnras}{7}{}{153}.
\newblock
\begin{APACrefDOI} \doi{10.1093/mnras/7.9.153} \end{APACrefDOI}
\PrintBackRefs{\CurrentBib}

\bibitem [\protect \citeauthoryear {%
Grafton-Waters%
\ \protect \BOthers {.}}{%
Grafton-Waters%
\ \protect \BOthers {.}}{%
{\protect \APACyear {2021}}%
}]{%
grafton_orbyts}
\APACinsertmetastar {%
grafton_orbyts}%
\begin{APACrefauthors}%
Grafton-Waters, S.%
, Ahmed, M.%
, Henson, S.%
, Hinds-Williams, F.%
, Ivanova, B.%
, Marshall, E.%
\BDBL {}Dunn, W.%
\end{APACrefauthors}%
\unskip\
\newblock
\APACrefYearMonthDay{2021}{jul}{}.
\newblock
{\BBOQ}\APACrefatitle {A Study of the Soft X-Ray Emission Lines in {NGC} 4151.
  I. Kinematic Properties of the Plasma Wind} {A study of the soft x-ray
  emission lines in {NGC} 4151. i. kinematic properties of the plasma
  wind}.{\BBCQ}
\newblock
\APACjournalVolNumPages{Research Notes of the {AAS}}{5}{7}{172}.
\newblock
\begin{APACrefURL} \url{https://doi.org/10.3847/2515-5172/ac1689}
  \end{APACrefURL}
\newblock
\begin{APACrefDOI} \doi{10.3847/2515-5172/ac1689} \end{APACrefDOI}
\PrintBackRefs{\CurrentBib}

\bibitem [\protect \citeauthoryear {%
{Hartman}%
\ \protect \BOthers {.}}{%
{Hartman}%
\ \protect \BOthers {.}}{%
{\protect \APACyear {2011}}%
}]{%
hartman_hatp18}
\APACinsertmetastar {%
hartman_hatp18}%
\begin{APACrefauthors}%
{Hartman}, J\BPBI D.%
, {Bakos}, G\BPBI {\'A}.%
, {Sato}, B.%
, {Torres}, G.%
, {Noyes}, R\BPBI W.%
, {Latham}, D\BPBI W.%
\BDBL {}{S{\'a}ri}, P.%
\end{APACrefauthors}%
\unskip\
\newblock
\APACrefYearMonthDay{2011}{{\APACmonth{01}}}{}.
\newblock
{\BBOQ}\APACrefatitle {{HAT-P-18b and HAT-P-19b: Two Low-density Saturn-mass
  Planets Transiting Metal-rich K Stars}} {{HAT-P-18b and HAT-P-19b: Two
  Low-density Saturn-mass Planets Transiting Metal-rich K Stars}}.{\BBCQ}
\newblock
\APACjournalVolNumPages{\apj}{726}{1}{52}.
\newblock
\begin{APACrefDOI} \doi{10.1088/0004-637X/726/1/52} \end{APACrefDOI}
\PrintBackRefs{\CurrentBib}

\bibitem [\protect \citeauthoryear {%
Hatzes%
\ \protect \BOthers {.}}{%
Hatzes%
\ \protect \BOthers {.}}{%
{\protect \APACyear {2003}}%
}]{%
hatzes_2003}
\APACinsertmetastar {%
hatzes_2003}%
\begin{APACrefauthors}%
Hatzes, A\BPBI P.%
, Cochran, W\BPBI D.%
, Endl, M.%
, McArthur, B.%
, Paulson, D\BPBI B.%
, Walker, G\BPBI A\BPBI H.%
\BDBL {}Yang, S.%
\end{APACrefauthors}%
\unskip\
\newblock
\APACrefYearMonthDay{2003}{dec}{}.
\newblock
{\BBOQ}\APACrefatitle {A Planetary Companion to Gamma Cephei A} {A planetary
  companion to gamma cephei a}.{\BBCQ}
\newblock
\APACjournalVolNumPages{The Astrophysical Journal}{599}{2}{1383--1394}.
\newblock
\begin{APACrefURL} \url{https://doi.org/10.1086/379281} \end{APACrefURL}
\newblock
\begin{APACrefDOI} \doi{10.1086/379281} \end{APACrefDOI}
\PrintBackRefs{\CurrentBib}

\bibitem [\protect \citeauthoryear {%
{Hellier}%
\ \protect \BOthers {.}}{%
{Hellier}%
\ \protect \BOthers {.}}{%
{\protect \APACyear {2012}}%
}]{%
hellier_w61}
\APACinsertmetastar {%
hellier_w61}%
\begin{APACrefauthors}%
{Hellier}, C.%
, {Anderson}, D\BPBI R.%
, {Collier Cameron}, A.%
, {Doyle}, A\BPBI P.%
, {Fumel}, A.%
, {Gillon}, M.%
\BDBL {}{West}, R\BPBI G.%
\end{APACrefauthors}%
\unskip\
\newblock
\APACrefYearMonthDay{2012}{{\APACmonth{10}}}{}.
\newblock
{\BBOQ}\APACrefatitle {{Seven transiting hot Jupiters from WASP-South, Euler
  and TRAPPIST: WASP-47b, WASP-55b, WASP-61b, WASP-62b, WASP-63b, WASP-66b and
  WASP-67b}} {{Seven transiting hot Jupiters from WASP-South, Euler and
  TRAPPIST: WASP-47b, WASP-55b, WASP-61b, WASP-62b, WASP-63b, WASP-66b and
  WASP-67b}}.{\BBCQ}
\newblock
\APACjournalVolNumPages{\mnras}{426}{1}{739-750}.
\newblock
\begin{APACrefDOI} \doi{10.1111/j.1365-2966.2012.21780.x} \end{APACrefDOI}
\PrintBackRefs{\CurrentBib}

\bibitem [\protect \citeauthoryear {%
Henry%
, Marcy%
, Butler%
\BCBL {}\ \BBA {} Vogt%
}{%
Henry%
\ \protect \BOthers {.}}{%
{\protect \APACyear {2000}}%
}]{%
Henry_2000}
\APACinsertmetastar {%
Henry_2000}%
\begin{APACrefauthors}%
Henry, G\BPBI W.%
, Marcy, G\BPBI W.%
, Butler, R\BPBI P.%
\BCBL {}\ \BBA {} Vogt, S\BPBI S.%
\end{APACrefauthors}%
\unskip\
\newblock
\APACrefYearMonthDay{2000}{jan}{}.
\newblock
{\BBOQ}\APACrefatitle {A Transiting {\textquotedblleft}51
  Peg{\textendash}like{\textquotedblright} Planet} {A transiting
  {\textquotedblleft}51 peg{\textendash}like{\textquotedblright}
  planet}.{\BBCQ}
\newblock
\APACjournalVolNumPages{The Astrophysical Journal}{529}{1}{L41--L44}.
\newblock
\begin{APACrefURL} \url{https://doi.org/10.1086/312458} \end{APACrefURL}
\newblock
\begin{APACrefDOI} \doi{10.1086/312458} \end{APACrefDOI}
\PrintBackRefs{\CurrentBib}

\bibitem [\protect \citeauthoryear {%
{Holdship}%
\ \protect \BOthers {.}}{%
{Holdship}%
\ \protect \BOthers {.}}{%
{\protect \APACyear {2019}}%
}]{%
holdship_orbyts}
\APACinsertmetastar {%
holdship_orbyts}%
\begin{APACrefauthors}%
{Holdship}, J.%
, {Viti}, S.%
, {Codella}, C.%
, {Rawlings}, J.%
, {Jimenez-Serra}, I.%
, {Ayalew}, Y.%
\BDBL {}{Horn}, S.%
\end{APACrefauthors}%
\unskip\
\newblock
\APACrefYearMonthDay{2019}{{\APACmonth{08}}}{}.
\newblock
{\BBOQ}\APACrefatitle {{Observations of CH$_{3}$OH and CH$_{3}$CHO in a Sample
  of Protostellar Outflow Sources}} {{Observations of CH$_{3}$OH and
  CH$_{3}$CHO in a Sample of Protostellar Outflow Sources}}.{\BBCQ}
\newblock
\APACjournalVolNumPages{\apj}{880}{2}{138}.
\newblock
\begin{APACrefDOI} \doi{10.3847/1538-4357/ab1f8f} \end{APACrefDOI}
\PrintBackRefs{\CurrentBib}

\bibitem [\protect \citeauthoryear {%
Hunter%
}{%
Hunter%
}{%
{\protect \APACyear {2007}}%
}]{%
Hunter_matplotlib}
\APACinsertmetastar {%
Hunter_matplotlib}%
\begin{APACrefauthors}%
Hunter, J\BPBI D.%
\end{APACrefauthors}%
\unskip\
\newblock
\APACrefYearMonthDay{2007}{}{}.
\newblock
{\BBOQ}\APACrefatitle {Matplotlib: A 2D graphics environment} {Matplotlib: A 2d
  graphics environment}.{\BBCQ}
\newblock
\APACjournalVolNumPages{Computing in Science \& Engineering}{9}{3}{90--95}.
\newblock
\begin{APACrefDOI} \doi{10.1109/MCSE.2007.55} \end{APACrefDOI}
\PrintBackRefs{\CurrentBib}

\bibitem [\protect \citeauthoryear {%
{Jacob}%
}{%
{Jacob}%
}{%
{\protect \APACyear {1855}}%
}]{%
jacob_1855}
\APACinsertmetastar {%
jacob_1855}%
\begin{APACrefauthors}%
{Jacob}, W\BPBI S.%
\end{APACrefauthors}%
\unskip\
\newblock
\APACrefYearMonthDay{1855}{{\APACmonth{06}}}{}.
\newblock
{\BBOQ}\APACrefatitle {{On certain Anomalies presented by the Binary Star 70
  Ophiuchi}} {{On certain Anomalies presented by the Binary Star 70
  Ophiuchi}}.{\BBCQ}
\newblock
\APACjournalVolNumPages{\mnras}{15}{}{228}.
\newblock
\begin{APACrefDOI} \doi{10.1093/mnras/15.9.228} \end{APACrefDOI}
\PrintBackRefs{\CurrentBib}

\bibitem [\protect \citeauthoryear {%
{Johnson}%
\ \protect \BOthers {.}}{%
{Johnson}%
\ \protect \BOthers {.}}{%
{\protect \APACyear {2011}}%
}]{%
johnson_h30}
\APACinsertmetastar {%
johnson_h30}%
\begin{APACrefauthors}%
{Johnson}, J\BPBI A.%
, {Winn}, J\BPBI N.%
, {Bakos}, G\BPBI {\'A}.%
, {Hartman}, J\BPBI D.%
, {Morton}, T\BPBI D.%
, {Torres}, G.%
\BDBL {}{S{\'a}ri}, P.%
\end{APACrefauthors}%
\unskip\
\newblock
\APACrefYearMonthDay{2011}{{\APACmonth{07}}}{}.
\newblock
{\BBOQ}\APACrefatitle {{HAT-P-30b: A Transiting Hot Jupiter on a Highly Oblique
  Orbit}} {{HAT-P-30b: A Transiting Hot Jupiter on a Highly Oblique
  Orbit}}.{\BBCQ}
\newblock
\APACjournalVolNumPages{\apj}{735}{1}{24}.
\newblock
\begin{APACrefDOI} \doi{10.1088/0004-637X/735/1/24} \end{APACrefDOI}
\PrintBackRefs{\CurrentBib}

\bibitem [\protect \citeauthoryear {%
{Juncher}%
\ \protect \BOthers {.}}{%
{Juncher}%
\ \protect \BOthers {.}}{%
{\protect \APACyear {2015}}%
}]{%
juncher_h55}
\APACinsertmetastar {%
juncher_h55}%
\begin{APACrefauthors}%
{Juncher}, D.%
, {Buchhave}, L\BPBI A.%
, {Hartman}, J\BPBI D.%
, {Bakos}, G\BPBI {\'A}.%
, {Bieryla}, A.%
, {Kov{\'a}cs}, T.%
\BDBL {}{S{\'a}ri}, P.%
\end{APACrefauthors}%
\unskip\
\newblock
\APACrefYearMonthDay{2015}{{\APACmonth{09}}}{}.
\newblock
{\BBOQ}\APACrefatitle {{HAT-P-55b: A Hot Jupiter Transiting a Sun-Like Star}}
  {{HAT-P-55b: A Hot Jupiter Transiting a Sun-Like Star}}.{\BBCQ}
\newblock
\APACjournalVolNumPages{\pasp}{127}{955}{851}.
\newblock
\begin{APACrefDOI} \doi{10.1086/682725} \end{APACrefDOI}
\PrintBackRefs{\CurrentBib}

\bibitem [\protect \citeauthoryear {%
{Kirk}%
\ \protect \BOthers {.}}{%
{Kirk}%
\ \protect \BOthers {.}}{%
{\protect \APACyear {2017}}%
}]{%
kirk_h18}
\APACinsertmetastar {%
kirk_h18}%
\begin{APACrefauthors}%
{Kirk}, J.%
, {Wheatley}, P\BPBI J.%
, {Louden}, T.%
, {Doyle}, A\BPBI P.%
, {Skillen}, I.%
, {McCormac}, J.%
\BDBL {}{Karjalainen}, R.%
\end{APACrefauthors}%
\unskip\
\newblock
\APACrefYearMonthDay{2017}{{\APACmonth{07}}}{}.
\newblock
{\BBOQ}\APACrefatitle {{Rayleigh scattering in the transmission spectrum of
  HAT-P-18b}} {{Rayleigh scattering in the transmission spectrum of
  HAT-P-18b}}.{\BBCQ}
\newblock
\APACjournalVolNumPages{\mnras}{468}{4}{3907-3916}.
\newblock
\begin{APACrefDOI} \doi{10.1093/mnras/stx752} \end{APACrefDOI}
\PrintBackRefs{\CurrentBib}

\bibitem [\protect \citeauthoryear {%
{Kokori}%
\ \protect \BOthers {.}}{%
{Kokori}%
\ \protect \BOthers {.}}{%
{\protect \APACyear {2021}}%
}]{%
kokori_exoclock}
\APACinsertmetastar {%
kokori_exoclock}%
\begin{APACrefauthors}%
{Kokori}, A.%
, {Tsiaras}, A.%
, {Edwards}, B.%
, {Rocchetto}, M.%
, {Tinetti}, G.%
, {W{\"u}nsche}, A.%
\BDBL {}{Tomatis}, A.%
\end{APACrefauthors}%
\unskip\
\newblock
\APACrefYearMonthDay{2021}{{\APACmonth{08}}}{}.
\newblock
{\BBOQ}\APACrefatitle {{ExoClock project: an open platform for monitoring the
  ephemerides of Ariel targets with contributions from the public}} {{ExoClock
  project: an open platform for monitoring the ephemerides of Ariel targets
  with contributions from the public}}.{\BBCQ}
\newblock
\APACjournalVolNumPages{Experimental Astronomy}{}{}{}.
\newblock
\begin{APACrefDOI} \doi{10.1007/s10686-020-09696-3} \end{APACrefDOI}
\PrintBackRefs{\CurrentBib}

\bibitem [\protect \citeauthoryear {%
{Le Verrier}%
}{%
{Le Verrier}%
}{%
{\protect \APACyear {1846}}%
}]{%
le_verrier_neptune}
\APACinsertmetastar {%
le_verrier_neptune}%
\begin{APACrefauthors}%
{Le Verrier}, U\BPBI J.%
\end{APACrefauthors}%
\unskip\
\newblock
\APACrefYearMonthDay{1846}{{\APACmonth{10}}}{}.
\newblock
{\BBOQ}\APACrefatitle {{Recherches sur les mouvements d'Uranus par U. J. Le
  Verrier (Fortsetzung).}} {{Recherches sur les mouvements d'Uranus par U. J.
  Le Verrier (Fortsetzung).}}{\BBCQ}
\newblock
\APACjournalVolNumPages{Astronomische Nachrichten}{25}{}{65}.
\PrintBackRefs{\CurrentBib}

\bibitem [\protect \citeauthoryear {%
{Lendl}%
\ \protect \BOthers {.}}{%
{Lendl}%
\ \protect \BOthers {.}}{%
{\protect \APACyear {2012}}%
}]{%
Lendl_WASP-42}
\APACinsertmetastar {%
Lendl_WASP-42}%
\begin{APACrefauthors}%
{Lendl}, M.%
, {Anderson}, D\BPBI R.%
, {Collier-Cameron}, A.%
, {Doyle}, A\BPBI P.%
, {Gillon}, M.%
, {Hellier}, C.%
\BDBL {}{Wheatley}, P\BPBI J.%
\end{APACrefauthors}%
\unskip\
\newblock
\APACrefYearMonthDay{2012}{{\APACmonth{08}}}{}.
\newblock
{\BBOQ}\APACrefatitle {{WASP-42 b and WASP-49 b: two new transiting
  sub-Jupiters}} {{WASP-42 b and WASP-49 b: two new transiting
  sub-Jupiters}}.{\BBCQ}
\newblock
\APACjournalVolNumPages{\aap}{544}{}{A72}.
\newblock
\begin{APACrefDOI} \doi{10.1051/0004-6361/201219585} \end{APACrefDOI}
\PrintBackRefs{\CurrentBib}

\bibitem [\protect \citeauthoryear {%
{Maciejewski}%
\ \protect \BOthers {.}}{%
{Maciejewski}%
\ \protect \BOthers {.}}{%
{\protect \APACyear {2016}}%
}]{%
mac_h30}
\APACinsertmetastar {%
mac_h30}%
\begin{APACrefauthors}%
{Maciejewski}, G.%
, {Dimitrov}, D.%
, {Mancini}, L.%
, {Southworth}, J.%
, {Ciceri}, S.%
, {D'Ago}, G.%
\BDBL {}{Henning}, T.%
\end{APACrefauthors}%
\unskip\
\newblock
\APACrefYearMonthDay{2016}{{\APACmonth{01}}}{}.
\newblock
{\BBOQ}\APACrefatitle {{New Transit Observations for HAT-P-30 b, HAT-P-37 b,
  TrES-5 b, WASP-28 b, WASP-36 b and WASP-39 b}} {{New Transit Observations for
  HAT-P-30 b, HAT-P-37 b, TrES-5 b, WASP-28 b, WASP-36 b and WASP-39
  b}}.{\BBCQ}
\newblock
\APACjournalVolNumPages{\actaa}{66}{1}{55-74}.
\PrintBackRefs{\CurrentBib}

\bibitem [\protect \citeauthoryear {%
{Mallonn}%
\ \protect \BOthers {.}}{%
{Mallonn}%
\ \protect \BOthers {.}}{%
{\protect \APACyear {2019}}%
}]{%
mallonn_ephm}
\APACinsertmetastar {%
mallonn_ephm}%
\begin{APACrefauthors}%
{Mallonn}, M.%
, {von Essen}, C.%
, {Herrero}, E.%
, {Alexoudi}, X.%
, {Granzer}, T.%
, {Sosa}, M.%
\BDBL {}{W{\"u}nsche}, A.%
\end{APACrefauthors}%
\unskip\
\newblock
\APACrefYearMonthDay{2019}{{\APACmonth{02}}}{}.
\newblock
{\BBOQ}\APACrefatitle {{Ephemeris refinement of 21 hot Jupiter exoplanets with
  high timing uncertainties}} {{Ephemeris refinement of 21 hot Jupiter
  exoplanets with high timing uncertainties}}.{\BBCQ}
\newblock
\APACjournalVolNumPages{\aap}{622}{}{A81}.
\newblock
\begin{APACrefDOI} \doi{10.1051/0004-6361/201834194} \end{APACrefDOI}
\PrintBackRefs{\CurrentBib}

\bibitem [\protect \citeauthoryear {%
Mayor%
\ \BBA {} Queloz%
}{%
Mayor%
\ \BBA {} Queloz%
}{%
{\protect \APACyear {1995}}%
}]{%
mayor_51peg}
\APACinsertmetastar {%
mayor_51peg}%
\begin{APACrefauthors}%
Mayor, M.%
\BCBT {}\ \BBA {} Queloz, D.%
\end{APACrefauthors}%
\unskip\
\newblock
\APACrefYearMonthDay{1995}{}{}.
\newblock
{\BBOQ}\APACrefatitle {A Jupiter-mass companion to a solar-type star} {A
  jupiter-mass companion to a solar-type star}.{\BBCQ}
\newblock
\APACjournalVolNumPages{Nature}{378}{6555}{355--359}.
\newblock
\begin{APACrefURL} \url{https://doi.org/10.1038/378355a0} \end{APACrefURL}
\newblock
\begin{APACrefDOI} \doi{10.1038/378355a0} \end{APACrefDOI}
\PrintBackRefs{\CurrentBib}

\bibitem [\protect \citeauthoryear {%
{McKemmish}%
\ \protect \BOthers {.}}{%
{McKemmish}%
\ \protect \BOthers {.}}{%
{\protect \APACyear {2018}}%
}]{%
mckemmish_orbytsII}
\APACinsertmetastar {%
mckemmish_orbytsII}%
\begin{APACrefauthors}%
{McKemmish}, L\BPBI K.%
, {Borsovszky}, J.%
, {Goodhew}, K\BPBI L.%
, {Sheppard}, S.%
, {Bennett}, A\BPBI F\BPBI V.%
, {Martin}, A\BPBI D\BPBI J.%
\BDBL {}{Tennyson}, J.%
\end{APACrefauthors}%
\unskip\
\newblock
\APACrefYearMonthDay{2018}{{\APACmonth{11}}}{}.
\newblock
{\BBOQ}\APACrefatitle {{MARVEL Analysis of the Measured High-resolution
  Rovibronic Spectra of $^{90}$Zr$^{16}$O}} {{MARVEL Analysis of the Measured
  High-resolution Rovibronic Spectra of $^{90}$Zr$^{16}$O}}.{\BBCQ}
\newblock
\APACjournalVolNumPages{\apj}{867}{1}{33}.
\newblock
\begin{APACrefDOI} \doi{10.3847/1538-4357/aadd19} \end{APACrefDOI}
\PrintBackRefs{\CurrentBib}

\bibitem [\protect \citeauthoryear {%
{McKemmish}%
\ \protect \BOthers {.}}{%
{McKemmish}%
\ \protect \BOthers {.}}{%
{\protect \APACyear {2017}}%
}]{%
mckemmish_orbyts}
\APACinsertmetastar {%
mckemmish_orbyts}%
\begin{APACrefauthors}%
{McKemmish}, L\BPBI K.%
, {Masseron}, T.%
, {Sheppard}, S.%
, {Sandeman}, E.%
, {Schofield}, Z.%
, {Furtenbacher}, T.%
\BDBL {}{Sousa-Silva}, C.%
\end{APACrefauthors}%
\unskip\
\newblock
\APACrefYearMonthDay{2017}{{\APACmonth{02}}}{}.
\newblock
{\BBOQ}\APACrefatitle {{Marvel Analysis of the Measured High-resolution
  Rovibronic Spectra of TiO}} {{Marvel Analysis of the Measured High-resolution
  Rovibronic Spectra of TiO}}.{\BBCQ}
\newblock
\APACjournalVolNumPages{\apjs}{228}{2}{15}.
\newblock
\begin{APACrefDOI} \doi{10.3847/1538-4365/228/2/15} \end{APACrefDOI}
\PrintBackRefs{\CurrentBib}

\bibitem [\protect \citeauthoryear {%
McKinney%
}{%
McKinney%
}{%
{\protect \APACyear {2011}}%
}]{%
mckinney_pandas}
\APACinsertmetastar {%
mckinney_pandas}%
\begin{APACrefauthors}%
McKinney, W.%
\end{APACrefauthors}%
\unskip\
\newblock
\APACrefYearMonthDay{2011}{}{}.
\newblock
{\BBOQ}\APACrefatitle {pandas: a foundational Python library for data analysis
  and statistics} {pandas: a foundational python library for data analysis and
  statistics}.{\BBCQ}
\newblock
\APACjournalVolNumPages{Python for High Performance and Scientific
  Computing}{14}{}{}.
\PrintBackRefs{\CurrentBib}

\bibitem [\protect \citeauthoryear {%
{Met Office}%
}{%
{Met Office}%
}{%
{\protect \APACyear {2010 - 2015}}%
}]{%
cartopy}
\APACinsertmetastar {%
cartopy}%
\begin{APACrefauthors}%
{Met Office}.%
\end{APACrefauthors}%
\unskip\
\newblock
\APACrefYearMonthDay{2010 - 2015}{}{}.
\newblock
{\BBOQ}\APACrefatitle {Cartopy: a cartographic python library with a matplotlib
  interface} {Cartopy: a cartographic python library with a matplotlib
  interface}{\BBCQ}\ [\bibcomputersoftwaremanual].
\newblock
\APACaddressPublisher{Exeter, Devon}{}.
\newblock
\begin{APACrefURL} \url{http://scitools.org.uk/cartopy} \end{APACrefURL}
\PrintBackRefs{\CurrentBib}

\bibitem [\protect \citeauthoryear {%
{Mohler-Fischer}%
\ \protect \BOthers {.}}{%
{Mohler-Fischer}%
\ \protect \BOthers {.}}{%
{\protect \APACyear {2013}}%
}]{%
mohler_h2}
\APACinsertmetastar {%
mohler_h2}%
\begin{APACrefauthors}%
{Mohler-Fischer}, M.%
, {Mancini}, L.%
, {Hartman}, J\BPBI D.%
, {Bakos}, G\BPBI {\'A}.%
, {Penev}, K.%
, {Bayliss}, D.%
\BDBL {}{Conroy}, P.%
\end{APACrefauthors}%
\unskip\
\newblock
\APACrefYearMonthDay{2013}{{\APACmonth{10}}}{}.
\newblock
{\BBOQ}\APACrefatitle {{HATS-2b: A transiting extrasolar planet orbiting a
  K-type star showing starspot activity}} {{HATS-2b: A transiting extrasolar
  planet orbiting a K-type star showing starspot activity}}.{\BBCQ}
\newblock
\APACjournalVolNumPages{\aap}{558}{}{A55}.
\newblock
\begin{APACrefDOI} \doi{10.1051/0004-6361/201321663} \end{APACrefDOI}
\PrintBackRefs{\CurrentBib}

\bibitem [\protect \citeauthoryear {%
{Morello}%
\ \protect \BOthers {.}}{%
{Morello}%
\ \protect \BOthers {.}}{%
{\protect \APACyear {2020}}%
}]{%
morello_exotethys}
\APACinsertmetastar {%
morello_exotethys}%
\begin{APACrefauthors}%
{Morello}, G.%
, {Claret}, A.%
, {Martin-Lagarde}, M.%
, {Cossou}, C.%
, {Tsiaras}, A.%
\BCBL {}\ \BBA {} {Lagage}, P\BPBI O.%
\end{APACrefauthors}%
\unskip\
\newblock
\APACrefYearMonthDay{2020}{{\APACmonth{02}}}{}.
\newblock
{\BBOQ}\APACrefatitle {{The ExoTETHyS Package: Tools for Exoplanetary Transits
  around Host Stars}} {{The ExoTETHyS Package: Tools for Exoplanetary Transits
  around Host Stars}}.{\BBCQ}
\newblock
\APACjournalVolNumPages{\aj}{159}{2}{75}.
\newblock
\begin{APACrefDOI} \doi{10.3847/1538-3881/ab63dc} \end{APACrefDOI}
\PrintBackRefs{\CurrentBib}

\bibitem [\protect \citeauthoryear {%
Oliphant%
}{%
Oliphant%
}{%
{\protect \APACyear {2006}}%
}]{%
oliphant_numpy}
\APACinsertmetastar {%
oliphant_numpy}%
\begin{APACrefauthors}%
Oliphant, T\BPBI E.%
\end{APACrefauthors}%
\unskip\
\newblock
\APACrefYear{2006}.
\newblock
\APACrefbtitle {A guide to NumPy} {A guide to numpy}\ (\BVOL~1).
\newblock
\APACaddressPublisher{}{Trelgol Publishing USA}.
\PrintBackRefs{\CurrentBib}

\bibitem [\protect \citeauthoryear {%
{Paragas}%
\ \protect \BOthers {.}}{%
{Paragas}%
\ \protect \BOthers {.}}{%
{\protect \APACyear {2021}}%
}]{%
paragas_h18}
\APACinsertmetastar {%
paragas_h18}%
\begin{APACrefauthors}%
{Paragas}, K.%
, {Vissapragada}, S.%
, {Knutson}, H\BPBI A.%
, {Oklop{\v{c}}i{\'c}}, A.%
, {Chachan}, Y.%
, {Greklek-McKeon}, M.%
\BDBL {}{Vasisht}, G.%
\end{APACrefauthors}%
\unskip\
\newblock
\APACrefYearMonthDay{2021}{{\APACmonth{03}}}{}.
\newblock
{\BBOQ}\APACrefatitle {{Metastable Helium Reveals an Extended Atmosphere for
  the Gas Giant HAT-P-18b}} {{Metastable Helium Reveals an Extended Atmosphere
  for the Gas Giant HAT-P-18b}}.{\BBCQ}
\newblock
\APACjournalVolNumPages{\apjl}{909}{1}{L10}.
\newblock
\begin{APACrefDOI} \doi{10.3847/2041-8213/abe706} \end{APACrefDOI}
\PrintBackRefs{\CurrentBib}

\bibitem [\protect \citeauthoryear {%
{Penev}%
\ \protect \BOthers {.}}{%
{Penev}%
\ \protect \BOthers {.}}{%
{\protect \APACyear {2013}}%
}]{%
penev_h1}
\APACinsertmetastar {%
penev_h1}%
\begin{APACrefauthors}%
{Penev}, K.%
, {Bakos}, G\BPBI {\'A}.%
, {Bayliss}, D.%
, {Jord{\'a}n}, A.%
, {Mohler}, M.%
, {Zhou}, G.%
\BDBL {}{S{\'a}ri}, P.%
\end{APACrefauthors}%
\unskip\
\newblock
\APACrefYearMonthDay{2013}{{\APACmonth{01}}}{}.
\newblock
{\BBOQ}\APACrefatitle {{HATS-1b: The First Transiting Planet Discovered by the
  HATSouth Survey}} {{HATS-1b: The First Transiting Planet Discovered by the
  HATSouth Survey}}.{\BBCQ}
\newblock
\APACjournalVolNumPages{\aj}{145}{1}{5}.
\newblock
\begin{APACrefDOI} \doi{10.1088/0004-6256/145/1/5} \end{APACrefDOI}
\PrintBackRefs{\CurrentBib}

\bibitem [\protect \citeauthoryear {%
{Poddan{\'y}}%
, {Br{\'a}t}%
\BCBL {}\ \BBA {} {Pejcha}%
}{%
{Poddan{\'y}}%
\ \protect \BOthers {.}}{%
{\protect \APACyear {2010}}%
}]{%
poddany_etd}
\APACinsertmetastar {%
poddany_etd}%
\begin{APACrefauthors}%
{Poddan{\'y}}, S.%
, {Br{\'a}t}, L.%
\BCBL {}\ \BBA {} {Pejcha}, O.%
\end{APACrefauthors}%
\unskip\
\newblock
\APACrefYearMonthDay{2010}{{\APACmonth{03}}}{}.
\newblock
{\BBOQ}\APACrefatitle {{Exoplanet Transit Database. Reduction and processing of
  the photometric data of exoplanet transits}} {{Exoplanet Transit Database.
  Reduction and processing of the photometric data of exoplanet
  transits}}.{\BBCQ}
\newblock
\APACjournalVolNumPages{\na}{15}{3}{297-301}.
\newblock
\begin{APACrefDOI} \doi{10.1016/j.newast.2009.09.001} \end{APACrefDOI}
\PrintBackRefs{\CurrentBib}

\bibitem [\protect \citeauthoryear {%
{Ricker}%
\ \protect \BOthers {.}}{%
{Ricker}%
\ \protect \BOthers {.}}{%
{\protect \APACyear {2015}}%
}]{%
ricker}
\APACinsertmetastar {%
ricker}%
\begin{APACrefauthors}%
{Ricker}, G\BPBI R.%
, {Winn}, J\BPBI N.%
, {Vanderspek}, R.%
, {Latham}, D\BPBI W.%
, {Bakos}, G\BPBI {\'A}.%
, {Bean}, J\BPBI L.%
\BDBL {}{Villasenor}, J.%
\end{APACrefauthors}%
\unskip\
\newblock
\APACrefYearMonthDay{2015}{{\APACmonth{01}}}{}.
\newblock
{\BBOQ}\APACrefatitle {{Transiting Exoplanet Survey Satellite (TESS)}}
  {{Transiting Exoplanet Survey Satellite (TESS)}}.{\BBCQ}
\newblock
\APACjournalVolNumPages{Journal of Astronomical Telescopes, Instruments, and
  Systems}{1}{}{014003}.
\newblock
\begin{APACrefDOI} \doi{10.1117/1.JATIS.1.1.014003} \end{APACrefDOI}
\PrintBackRefs{\CurrentBib}

\bibitem [\protect \citeauthoryear {%
{See}%
}{%
{See}%
}{%
{\protect \APACyear {1896}}%
}]{%
see_1896}
\APACinsertmetastar {%
see_1896}%
\begin{APACrefauthors}%
{See}, T\BPBI J\BPBI J.%
\end{APACrefauthors}%
\unskip\
\newblock
\APACrefYearMonthDay{1896}{{\APACmonth{01}}}{}.
\newblock
{\BBOQ}\APACrefatitle {{Researches on the orbit of 70 Ophiuchi, and on a
  periodic perturbation in the motion of the system arising from the action of
  an unseen body}} {{Researches on the orbit of 70 Ophiuchi, and on a periodic
  perturbation in the motion of the system arising from the action of an unseen
  body}}.{\BBCQ}
\newblock
\APACjournalVolNumPages{\aj}{16}{}{17-23}.
\newblock
\begin{APACrefDOI} \doi{10.1086/102368} \end{APACrefDOI}
\PrintBackRefs{\CurrentBib}

\bibitem [\protect \citeauthoryear {%
{Seeliger}%
\ \protect \BOthers {.}}{%
{Seeliger}%
\ \protect \BOthers {.}}{%
{\protect \APACyear {2015}}%
}]{%
seeliger_h18_h27}
\APACinsertmetastar {%
seeliger_h18_h27}%
\begin{APACrefauthors}%
{Seeliger}, M.%
, {Kitze}, M.%
, {Errmann}, R.%
, {Richter}, S.%
, {Ohlert}, J\BPBI M.%
, {Chen}, W\BPBI P.%
\BDBL {}{Neuh{\"a}user}, R.%
\end{APACrefauthors}%
\unskip\
\newblock
\APACrefYearMonthDay{2015}{{\APACmonth{08}}}{}.
\newblock
{\BBOQ}\APACrefatitle {{Ground-based transit observations of the HAT-P-18,
  HAT-P-19, HAT-P-27/WASP40 and WASP-21 systems}} {{Ground-based transit
  observations of the HAT-P-18, HAT-P-19, HAT-P-27/WASP40 and WASP-21
  systems}}.{\BBCQ}
\newblock
\APACjournalVolNumPages{\mnras}{451}{4}{4060-4072}.
\newblock
\begin{APACrefDOI} \doi{10.1093/mnras/stv1187} \end{APACrefDOI}
\PrintBackRefs{\CurrentBib}

\bibitem [\protect \citeauthoryear {%
{Sherrill}%
}{%
{Sherrill}%
}{%
{\protect \APACyear {1999}}%
}]{%
sherrill}
\APACinsertmetastar {%
sherrill}%
\begin{APACrefauthors}%
{Sherrill}, T\BPBI J.%
\end{APACrefauthors}%
\unskip\
\newblock
\APACrefYearMonthDay{1999}{{\APACmonth{02}}}{}.
\newblock
{\BBOQ}\APACrefatitle {{A Career of Controversy: The Anomaly of T. J. J. See}}
  {{A Career of Controversy: The Anomaly of T. J. J. See}}.{\BBCQ}
\newblock
\APACjournalVolNumPages{Journal for the History of Astronomy}{30}{}{25}.
\newblock
\begin{APACrefDOI} \doi{10.1177/002182869903000102} \end{APACrefDOI}
\PrintBackRefs{\CurrentBib}

\bibitem [\protect \citeauthoryear {%
{Sousa-Silva}%
\ \protect \BOthers {.}}{%
{Sousa-Silva}%
\ \protect \BOthers {.}}{%
{\protect \APACyear {2018}}%
}]{%
sousa_silva_orbyts}
\APACinsertmetastar {%
sousa_silva_orbyts}%
\begin{APACrefauthors}%
{Sousa-Silva}, C.%
, {McKemmish}, L\BPBI K.%
, {Chubb}, K\BPBI L.%
, {Gorman}, M\BPBI N.%
, {Baker}, J\BPBI S.%
, {Barton}, E\BPBI J.%
\BDBL {}{Tennyson}, J.%
\end{APACrefauthors}%
\unskip\
\newblock
\APACrefYearMonthDay{2018}{{\APACmonth{01}}}{}.
\newblock
{\BBOQ}\APACrefatitle {{Original Research By Young Twinkle Students (ORBYTS):
  when can students start performing original research?}} {{Original Research
  By Young Twinkle Students (ORBYTS): when can students start performing
  original research?}}{\BBCQ}
\newblock
\APACjournalVolNumPages{Physics Education}{53}{1}{015020}.
\newblock
\begin{APACrefDOI} \doi{10.1088/1361-6552/aa8f2a} \end{APACrefDOI}
\PrintBackRefs{\CurrentBib}

\bibitem [\protect \citeauthoryear {%
{Southworth}%
\ \protect \BOthers {.}}{%
{Southworth}%
\ \protect \BOthers {.}}{%
{\protect \APACyear {2014}}%
}]{%
southworth_w25}
\APACinsertmetastar {%
southworth_w25}%
\begin{APACrefauthors}%
{Southworth}, J.%
, {Hinse}, T\BPBI C.%
, {Burgdorf}, M.%
, {Calchi Novati}, S.%
, {Dominik}, M.%
, {Galianni}, P.%
\BDBL {}{Vilela}, C.%
\end{APACrefauthors}%
\unskip\
\newblock
\APACrefYearMonthDay{2014}{{\APACmonth{10}}}{}.
\newblock
{\BBOQ}\APACrefatitle {{High-precision photometry by telescope defocussing -
  VI. WASP-24, WASP-25 and WASP-26}} {{High-precision photometry by telescope
  defocussing - VI. WASP-24, WASP-25 and WASP-26}}.{\BBCQ}
\newblock
\APACjournalVolNumPages{\mnras}{444}{1}{776-789}.
\newblock
\begin{APACrefDOI} \doi{10.1093/mnras/stu1492} \end{APACrefDOI}
\PrintBackRefs{\CurrentBib}

\bibitem [\protect \citeauthoryear {%
{Southworth}%
\ \protect \BOthers {.}}{%
{Southworth}%
\ \protect \BOthers {.}}{%
{\protect \APACyear {2015}}%
}]{%
Southworth_WASP-57}
\APACinsertmetastar {%
Southworth_WASP-57}%
\begin{APACrefauthors}%
{Southworth}, J.%
, {Mancini}, L.%
, {Tregloan-Reed}, J.%
, {Calchi Novati}, S.%
, {Ciceri}, S.%
, {D'Ago}, G.%
\BDBL {}{Wertz}, O.%
\end{APACrefauthors}%
\unskip\
\newblock
\APACrefYearMonthDay{2015}{{\APACmonth{12}}}{}.
\newblock
{\BBOQ}\APACrefatitle {{Larger and faster: revised properties and a shorter
  orbital period for the WASP-57 planetary system from a pro-am collaboration}}
  {{Larger and faster: revised properties and a shorter orbital period for the
  WASP-57 planetary system from a pro-am collaboration}}.{\BBCQ}
\newblock
\APACjournalVolNumPages{\mnras}{454}{3}{3094-3107}.
\newblock
\begin{APACrefDOI} \doi{10.1093/mnras/stv2183} \end{APACrefDOI}
\PrintBackRefs{\CurrentBib}

\bibitem [\protect \citeauthoryear {%
{Southworth}%
\ \protect \BOthers {.}}{%
{Southworth}%
\ \protect \BOthers {.}}{%
{\protect \APACyear {2016}}%
}]{%
southworth_w42}
\APACinsertmetastar {%
southworth_w42}%
\begin{APACrefauthors}%
{Southworth}, J.%
, {Tregloan-Reed}, J.%
, {Andersen}, M\BPBI I.%
, {Calchi Novati}, S.%
, {Ciceri}, S.%
, {Colque}, J\BPBI P.%
\BDBL {}{Wang}, Y.%
\end{APACrefauthors}%
\unskip\
\newblock
\APACrefYearMonthDay{2016}{{\APACmonth{04}}}{}.
\newblock
{\BBOQ}\APACrefatitle {{High-precision photometry by telescope defocussing -
  VIII. WASP-22, WASP-41, WASP-42 and WASP-55}} {{High-precision photometry by
  telescope defocussing - VIII. WASP-22, WASP-41, WASP-42 and WASP-55}}.{\BBCQ}
\newblock
\APACjournalVolNumPages{\mnras}{457}{4}{4205-4217}.
\newblock
\begin{APACrefDOI} \doi{10.1093/mnras/stw279} \end{APACrefDOI}
\PrintBackRefs{\CurrentBib}

\bibitem [\protect \citeauthoryear {%
{Struve}%
}{%
{Struve}%
}{%
{\protect \APACyear {1952}}%
}]{%
struve_1952}
\APACinsertmetastar {%
struve_1952}%
\begin{APACrefauthors}%
{Struve}, O.%
\end{APACrefauthors}%
\unskip\
\newblock
\APACrefYearMonthDay{1952}{{\APACmonth{10}}}{}.
\newblock
{\BBOQ}\APACrefatitle {{Proposal for a project of high-precision stellar radial
  velocity work}} {{Proposal for a project of high-precision stellar radial
  velocity work}}.{\BBCQ}
\newblock
\APACjournalVolNumPages{The Observatory}{72}{}{199-200}.
\PrintBackRefs{\CurrentBib}

\bibitem [\protect \citeauthoryear {%
{Tinetti}%
\ \protect \BOthers {.}}{%
{Tinetti}%
\ \protect \BOthers {.}}{%
{\protect \APACyear {2018}}%
}]{%
tinetti_ariel}
\APACinsertmetastar {%
tinetti_ariel}%
\begin{APACrefauthors}%
{Tinetti}, G.%
, {Drossart}, P.%
, {Eccleston}, P.%
, {Hartogh}, P.%
, {Heske}, A.%
, {Leconte}, J.%
\BDBL {}{Zwart}, F.%
\end{APACrefauthors}%
\unskip\
\newblock
\APACrefYearMonthDay{2018}{{\APACmonth{11}}}{}.
\newblock
{\BBOQ}\APACrefatitle {{A chemical survey of exoplanets with ARIEL}} {{A
  chemical survey of exoplanets with ARIEL}}.{\BBCQ}
\newblock
\APACjournalVolNumPages{Experimental Astronomy}{46}{1}{135-209}.
\newblock
\begin{APACrefDOI} \doi{10.1007/s10686-018-9598-x} \end{APACrefDOI}
\PrintBackRefs{\CurrentBib}

\bibitem [\protect \citeauthoryear {%
{Tinetti}%
\ \protect \BOthers {.}}{%
{Tinetti}%
\ \protect \BOthers {.}}{%
{\protect \APACyear {2021}}%
}]{%
tinetti_ariel2}
\APACinsertmetastar {%
tinetti_ariel2}%
\begin{APACrefauthors}%
{Tinetti}, G.%
, {Eccleston}, P.%
, {Haswell}, C.%
, {Lagage}, P\BHBI O.%
, {Leconte}, J.%
, {L{\"u}ftinger}, T.%
\BDBL {}{Zuppella}, P.%
\end{APACrefauthors}%
\unskip\
\newblock
\APACrefYearMonthDay{2021}{{\APACmonth{04}}}{}.
\newblock
{\BBOQ}\APACrefatitle {{Ariel: Enabling planetary science across light-years}}
  {{Ariel: Enabling planetary science across light-years}}.{\BBCQ}
\newblock
\APACjournalVolNumPages{arXiv e-prints}{}{}{arXiv:2104.04824}.
\PrintBackRefs{\CurrentBib}

\bibitem [\protect \citeauthoryear {%
{Tsiaras}%
}{%
{Tsiaras}%
}{%
{\protect \APACyear {2019}}%
}]{%
tsiaras_hops}
\APACinsertmetastar {%
tsiaras_hops}%
\begin{APACrefauthors}%
{Tsiaras}, A.%
\end{APACrefauthors}%
\unskip\
\newblock
\APACrefYearMonthDay{2019}{{\APACmonth{09}}}{}.
\newblock
{\BBOQ}\APACrefatitle {{HOPS: the photometric software of the HOlomon
  Astronomical Station}} {{HOPS: the photometric software of the HOlomon
  Astronomical Station}}.{\BBCQ}
\newblock
\BIn{} \APACrefbtitle {EPSC-DPS Joint Meeting 2019} {Epsc-dps joint meeting
  2019}\ (\BVOL\ 2019, \BPG~EPSC-DPS2019-1594).
\PrintBackRefs{\CurrentBib}

\bibitem [\protect \citeauthoryear {%
{Tsiaras}%
\ \protect \BOthers {.}}{%
{Tsiaras}%
\ \protect \BOthers {.}}{%
{\protect \APACyear {2016}}%
}]{%
tsiaras_plc}
\APACinsertmetastar {%
tsiaras_plc}%
\begin{APACrefauthors}%
{Tsiaras}, A.%
, {Waldmann}, I\BPBI P.%
, {Rocchetto}, M.%
, {Varley}, R.%
, {Morello}, G.%
, {Damiano}, M.%
\BCBL {}\ \BBA {} {Tinetti}, G.%
\end{APACrefauthors}%
\unskip\
\newblock
\APACrefYearMonthDay{2016}{{\APACmonth{12}}}{}.
\newblock
\APACrefbtitle {{pylightcurve: Exoplanet lightcurve model}.} {{pylightcurve:
  Exoplanet lightcurve model}.}
\PrintBackRefs{\CurrentBib}

\bibitem [\protect \citeauthoryear {%
{Tsiaras}%
\ \protect \BOthers {.}}{%
{Tsiaras}%
\ \protect \BOthers {.}}{%
{\protect \APACyear {2018}}%
}]{%
tsiaras_30planets}
\APACinsertmetastar {%
tsiaras_30planets}%
\begin{APACrefauthors}%
{Tsiaras}, A.%
, {Waldmann}, I\BPBI P.%
, {Zingales}, T.%
, {Rocchetto}, M.%
, {Morello}, G.%
, {Damiano}, M.%
\BDBL {}{Yurchenko}, S\BPBI N.%
\end{APACrefauthors}%
\unskip\
\newblock
\APACrefYearMonthDay{2018}{{\APACmonth{04}}}{}.
\newblock
{\BBOQ}\APACrefatitle {{A Population Study of Gaseous Exoplanets}} {{A
  Population Study of Gaseous Exoplanets}}.{\BBCQ}
\newblock
\APACjournalVolNumPages{\aj}{155}{4}{156}.
\newblock
\begin{APACrefDOI} \doi{10.3847/1538-3881/aaaf75} \end{APACrefDOI}
\PrintBackRefs{\CurrentBib}

\bibitem [\protect \citeauthoryear {%
{Turner}%
\ \protect \BOthers {.}}{%
{Turner}%
\ \protect \BOthers {.}}{%
{\protect \APACyear {2016}}%
}]{%
turner_w123}
\APACinsertmetastar {%
turner_w123}%
\begin{APACrefauthors}%
{Turner}, O\BPBI D.%
, {Anderson}, D\BPBI R.%
, {Collier Cameron}, A.%
, {Delrez}, L.%
, {Evans}, D\BPBI F.%
, {Gillon}, M.%
\BDBL {}{West}, R\BPBI G.%
\end{APACrefauthors}%
\unskip\
\newblock
\APACrefYearMonthDay{2016}{{\APACmonth{06}}}{}.
\newblock
{\BBOQ}\APACrefatitle {{WASP-120 b, WASP-122 b, AND WASP-123 b: Three Newly
  Discovered Planets from the WASP-South Survey}} {{WASP-120 b, WASP-122 b, AND
  WASP-123 b: Three Newly Discovered Planets from the WASP-South
  Survey}}.{\BBCQ}
\newblock
\APACjournalVolNumPages{\pasp}{128}{964}{064401}.
\newblock
\begin{APACrefDOI} \doi{10.1088/1538-3873/128/964/064401} \end{APACrefDOI}
\PrintBackRefs{\CurrentBib}

\bibitem [\protect \citeauthoryear {%
{van de Kamp}%
}{%
{van de Kamp}%
}{%
{\protect \APACyear {1969}}%
}]{%
van_kemp}
\APACinsertmetastar {%
van_kemp}%
\begin{APACrefauthors}%
{van de Kamp}, P.%
\end{APACrefauthors}%
\unskip\
\newblock
\APACrefYearMonthDay{1969}{{\APACmonth{08}}}{}.
\newblock
{\BBOQ}\APACrefatitle {{Alternate dynamical analysis of Barnard's star.}}
  {{Alternate dynamical analysis of Barnard's star.}}{\BBCQ}
\newblock
\APACjournalVolNumPages{\aj}{74}{}{757-759}.
\newblock
\begin{APACrefDOI} \doi{10.1086/110852} \end{APACrefDOI}
\PrintBackRefs{\CurrentBib}

\bibitem [\protect \citeauthoryear {%
{Virtanen}%
\ \protect \BOthers {.}}{%
{Virtanen}%
\ \protect \BOthers {.}}{%
{\protect \APACyear {2020}}%
}]{%
scipy}
\APACinsertmetastar {%
scipy}%
\begin{APACrefauthors}%
{Virtanen}, P.%
, {Gommers}, R.%
, {Oliphant}, T\BPBI E.%
, {Haberland}, M.%
, {Reddy}, T.%
, {Cournapeau}, D.%
\BDBL {}{Contributors}, S\BPBI \BPBI .%
\end{APACrefauthors}%
\unskip\
\newblock
\APACrefYearMonthDay{2020}{}{}.
\newblock
{\BBOQ}\APACrefatitle {{SciPy 1.0: Fundamental Algorithms for Scientific
  Computing in Python}} {{SciPy 1.0: Fundamental Algorithms for Scientific
  Computing in Python}}.{\BBCQ}
\newblock
\APACjournalVolNumPages{Nature Methods}{17}{}{261--272}.
\newblock
\begin{APACrefDOI} \doi{https://doi.org/10.1038/s41592-019-0686-2}
  \end{APACrefDOI}
\PrintBackRefs{\CurrentBib}

\bibitem [\protect \citeauthoryear {%
Wibisono%
\ \protect \BOthers {.}}{%
Wibisono%
\ \protect \BOthers {.}}{%
{\protect \APACyear {2020}}%
}]{%
wibisono_orbyts}
\APACinsertmetastar {%
wibisono_orbyts}%
\begin{APACrefauthors}%
Wibisono, A\BPBI D.%
, Branduardi-Raymont, G.%
, Dunn, W\BPBI R.%
, Coates, A\BPBI J.%
, Weigt, D\BPBI M.%
, Jackman, C\BPBI M.%
\BDBL {}Fleming, D.%
\end{APACrefauthors}%
\unskip\
\newblock
\APACrefYearMonthDay{2020}{}{}.
\newblock
{\BBOQ}\APACrefatitle {Temporal and Spectral Studies by XMM-Newton of Jupiter's
  X-ray Auroras During a Compression Event} {Temporal and spectral studies by
  xmm-newton of jupiter's x-ray auroras during a compression event}.{\BBCQ}
\newblock
\APACjournalVolNumPages{Journal of Geophysical Research: Space
  Physics}{125}{5}{e2019JA027676}.
\newblock
\begin{APACrefURL}
  \url{https://agupubs.onlinelibrary.wiley.com/doi/abs/10.1029/2019JA027676}
  \end{APACrefURL}
\newblock
\APACrefnote{e2019JA027676 10.1029/2019JA027676}
\newblock
\begin{APACrefDOI} \doi{https://doi.org/10.1029/2019JA027676} \end{APACrefDOI}
\PrintBackRefs{\CurrentBib}

\bibitem [\protect \citeauthoryear {%
{Wolszczan}%
\ \BBA {} {Frail}%
}{%
{Wolszczan}%
\ \BBA {} {Frail}%
}{%
{\protect \APACyear {1992}}%
}]{%
wolszczan_pulsar}
\APACinsertmetastar {%
wolszczan_pulsar}%
\begin{APACrefauthors}%
{Wolszczan}, A.%
\BCBT {}\ \BBA {} {Frail}, D\BPBI A.%
\end{APACrefauthors}%
\unskip\
\newblock
\APACrefYearMonthDay{1992}{{\APACmonth{01}}}{}.
\newblock
{\BBOQ}\APACrefatitle {{A planetary system around the millisecond pulsar
  PSR1257 + 12}} {{A planetary system around the millisecond pulsar PSR1257 +
  12}}.{\BBCQ}
\newblock
\APACjournalVolNumPages{\nat}{355}{6356}{145-147}.
\newblock
\begin{APACrefDOI} \doi{10.1038/355145a0} \end{APACrefDOI}
\PrintBackRefs{\CurrentBib}

\bibitem [\protect \citeauthoryear {%
{Zellem}%
\ \protect \BOthers {.}}{%
{Zellem}%
\ \protect \BOthers {.}}{%
{\protect \APACyear {2020}}%
}]{%
zellem_exowatch}
\APACinsertmetastar {%
zellem_exowatch}%
\begin{APACrefauthors}%
{Zellem}, R\BPBI T.%
, {Pearson}, K\BPBI A.%
, {Blaser}, E.%
, {Fowler}, M.%
, {Ciardi}, D\BPBI R.%
, {Biferno}, A.%
\BDBL {}{Malvache}, A.%
\end{APACrefauthors}%
\unskip\
\newblock
\APACrefYearMonthDay{2020}{{\APACmonth{05}}}{}.
\newblock
{\BBOQ}\APACrefatitle {{Utilizing Small Telescopes Operated by Citizen
  Scientists for Transiting Exoplanet Follow-up}} {{Utilizing Small Telescopes
  Operated by Citizen Scientists for Transiting Exoplanet Follow-up}}.{\BBCQ}
\newblock
\APACjournalVolNumPages{\pasp}{132}{1011}{054401}.
\newblock
\begin{APACrefDOI} \doi{10.1088/1538-3873/ab7ee7} \end{APACrefDOI}
\PrintBackRefs{\CurrentBib}

\end{thebibliography}
\bibliographystyle{apacite}

\end{document}